\documentclass[draft]{llncs}
\usepackage{amsfonts,amsmath,amssymb,color,latexsym,makeidx,times}
\definecolor{monblu}{cmyk}{0.80,0.00,0.00,0.70}
\definecolor{monmag}{cmyk}{0.00,1.00,0.00,0.00}
\def\N{\mathbb{N}}
\def\rulenecessitation{\mathbf{NEC}}
\def\reductionaxiom{\mathbf{RA}}
\def\modalaxiom{\mathbf{MA}}
\def\Fo{\mathbf{FOR}}
\def\modal{\mathbf{m}}
\def\direct{\mathbf{d}}
\def\simple{\mathbf{s}}
\def\reflexive{\mathbf{ref}}
\def\symmetric{\mathbf{sym}}
\def\partition{\mathbf{equ}}
\def\At{\mathbf{At}}
\def\K{\mathbf{K}}
\def\iK{\mathbf{iK}}
\def\IPAL{\mathbf{IPAL}}
\def\PAL{\mathbf{PAL}}
\def\IPL{\mathbf{IPL}}
\def\modusponens{\mathbf{MP}}
\def\L{\mathbf{L}}
\def\Sf{\mathtt{sf}}
\def\degre{\mathtt{deg}}
\begin{document}
\title{A minimal intuitionistic public announcement logic}
\author{Philippe Balbiani$^{\mathsection}$
\hspace{0.75cm}
Hans van Ditmarsch$^{\dagger,\mathsection}$
\hspace{0.75cm}
\\
\c{C}i\u gdem Gencer$^{\mathsection}$
\hspace{0.75cm}
Clara Lerouvillois$^{\ddagger,\mathsection}$}
\institute{
IIT Kanpur\footnote{Department of Computer Science and Engineering, Indian Institute of Technology Kanpur.}
\hspace{0.75cm}
IHPST\footnote{Institut d'histoire et de philosophie des sciences et des techniques, Universit\'e Paris 1 Panth\'eon-Sorbonne~---~CNRS.}
\hspace{0.75cm}
IRIT\footnote{Institut de recherche en informatique de Toulouse, CNRS~---~Toulouse INP~---~Universit\'e de Toulouse.}
}
\maketitle
\begin{abstract}
We introduce intuitionistic public announcement logics $\L_{\min}$ and
\linebreak$
\L_{\partition}$ determined by classes of birelational frames where the intuitionistic preorder and the modal accessibility relation do not need to satisfy the confluence conditions traditionally considered within the framework of $\IPAL$.
Our logics can be seen as generalisations of $\IPAL$ seeing that their language is based on a binary epistemic modality of ``relativised knowledge'' that allows us to define valid reduction axioms.
Our main results are the soundness and completeness of $\L_{\min}$ with respect to the class of all birelational frames and the soundness and completeness of $\L_{\partition}$ with respect to the class of all birelational frames where the modal accessibility relation is an equivalence relation.
\end{abstract}
\section{Introduction}
$\IPAL$~---~the intuitionistic public announcement logic introduced in~\cite{Ma:Palmigiano:Sadrzadeh:2014,Maffezioli:Negri:2011}~---~constitutes the subject of numerous discussions from the point of view of its model theory, the point of view of its proof theory, etc~\cite{Murai:Sano:2022,Nomura:Sano:Tojo:2015}.
In all these discussions, strong hypotheses concerning the birelational models $(W,{\leq},{R},V)$ in which the formulas are interpreted are made in order to be able to achieve four essential objectives.
%
%
%
%
The first and second objectives date back to~\cite{FischerServi:1984,Plotkin:Stirling:1986}.
They respectively consist of interpreting the modal operators $\square$ and $\lozenge$ of ``necessity'' and ``possibility'' in the way Kripke~\cite{Kripke:1965} interprets the first-order quantifiers $\forall$ and $\exists$ in his definition of the possible-worlds semantics of intuitionistic first-order logic.
That is to say, for all formulas $C$, in a birelational model $(W,{\leq},{R},V)$, given $s{\in}W$, ${\square}C$ holds at $s$ if for all $t{\in}W$, if $s{\leq}t$ then for all $u{\in}W$, if $t{R}u$ then $C$ holds at $u$ and ${\lozenge}C$ holds at $s$ if there exists $t{\in}W$ such that $s{R}t$ and $C$ holds at $t$.
The third objective is specific to logics based on public announcement operators~\cite{vanDitmarsch:et:al:2007}.
It concerns the existence of reduction axioms which make it possible to associate with each formula a semantically equivalent formula containing no occurrence of these operators.
For instance, the reduction axiom for atoms ${\lbrack A\rbrack}p{\leftrightarrow}(A{\rightarrow}p)$, the reduction axiom for ``false'' ${\lbrack A\rbrack}{\bot}{\leftrightarrow}{\neg}A$, etc.\footnote{Here, it is needed to mention how a public announcement operator is intuitionistically interpreted: for all formulas $C$, in a birelational model $(W,{\leq},{R},V)$, given $s{\in}W$, ${\lbrack A\rbrack}C$ holds at $s$ in $(W,{\leq},{R},V)$ if for all $t{\in}W$, if $s{\leq}t$ and $A$ holds at $t$ in $(W,{\leq},{R},V)$ then $C$ holds at $t$ in the restriction of $(W,{\leq},{R},V)$ to the set of all $u{\in}W$ such that $A$ holds at $u$ in $(W,{\leq},{R},V)$.
See~\cite{Balbiani:Galmiche:2016} for details.}
The fourth objective~---~probably the most important~---~is specific to logics based on intuitionist implication.
Indeed, it concerns the property of heredity which requires the truth sets associated to each formula to be increasing relative to the preorders which equip the birelational models.
That is to say, for all formulas $C$, in a birelational model $(W,{\leq},{R},V)$, given $s,t{\in}W$ such that $s{\leq}t$, if $C$ holds at $s$ then $C$ holds at $t$.
%
%
%
%
To achieve these objectives, all authors without exception choose to only interpret formulas in backward and forward confluent birelational models, i.e. birelational models $(W,{\leq},{R},V)$ such that ${R}{\circ}{\leq}{\subseteq}{\leq}{\circ}{R}$ and ${R^{-1}}{\circ}{\leq}{\subseteq}{\leq}{\circ}{R^{-1}}$, seeing that these hypotheses guarantee the validity of the reduction axiom for ``necessity'' ${\lbrack A\rbrack}{\square}C{\leftrightarrow}(A{\rightarrow}{\square}{\lbrack A\rbrack}C)$ and the property of heredity for $\lozenge$-formulas.
%
%
%
%
%
%
\\
\\
Today, we want to depart from the tradition by proposing a minimalist approach to $\IPAL$ for which no hypothesis about the structures underlying its model theory is made.
Unfortunately, that is easier said than done, seeing that without the hypotheses of backward and forward confluence, the reduction axiom for ``necessity'' ${\lbrack A\rbrack}{\square}C{\leftrightarrow}(A{\rightarrow}
$\linebreak$
{\square}{\lbrack A\rbrack}C)$ is no longer valid whereas the property of heredity for $\lozenge$-formu\-las is no longer verified.
Indeed, it does not seem possible to achieve our goal without modifying the language of $\IPAL$.\footnote{Indeed, without modifying the language of $\IPAL$, the question of whether a reduction axiom for ``necessity'' exists with respect to the class of all birelational models remains open.}
%
%
%
%
In this article, leaving aside for the moment all considerations regarding the operator $\lozenge$ of ``possibility'', we replace the operator $\square$ of ``necessity'' with a new binary operator $\triangleright$ of ``relativized necessity'' interpreted as follows in a birelational model $(W,{\leq},{R},V)$: given $s{\in}W$, $C{\triangleright}D$ holds at $s$ if for all $t{\in}W$, if $s{\leq}t$ then for all $u{\in}W$, if $t{R}u$ and $C$ holds at $u$ then $D$ holds at $u$.
%
%
%
%
The reader may easily verify that $\square$ is definable in terms of $\triangleright$, seeing that for all formulas $D$, ${\square}D$ holds exactly where ${\top}{\triangleright}D$ holds.
More importantly, a reduction axiom for ``relativized necessity'' can be proposed: the reduction axiom ${\lbrack A\rbrack}(C{\triangleright}D){\leftrightarrow}(A{\rightarrow}((A{\wedge}{\lbrack A\rbrack}C){\triangleright}{\lbrack A\rbrack}D))$ whose validity does not require the hypothesis of backward confluence.
Together with the modal axioms $(\modalaxiom1)$, $(\modalaxiom2)$ and $(\modalaxiom3)$ and the necessitation rule $(\rulenecessitation)$ presented in Section~\ref{section:axiomatization} and characterizing the properties of $\triangleright$, the reduction axiom for ``relativized necessity''~---~when added to the reduction axioms already known for atoms, for ``false'' , etc~---~makes it possible to axiomatize the validities determined by the class of all birelational models.
The addition in Section~\ref{section:axiomatization} of two other modal axioms $(\modalaxiom4)$ and $(\modalaxiom5)$ characterizing further properties of $\triangleright$ makes it possible to axiomatize the validities determined by the class of all birelational models where the modal accessibility relation is an equivalence relation.
%
%
%
\\
\\
In Sections~\ref{section:syntax} and~\ref{section:semantics}, we introduce the syntax and the semantics of our version of $\IPAL$.
In Section~\ref{section:axiomatization}, we present the definitions of our intuitionistic public announcement logics $\L_{\min}$ and $\L_{\partition}$.
In Section~\ref{section:theories}, we present results needed for the proofs developed in Section~\ref{section:completeness} of the completeness of $\L_{\min}$ with respect to validity in the class ${\mathcal C}_{\min}$ of all birelational models and the completeness of $\L_{\partition}$ with respect to validity in the class ${\mathcal C}_{\partition}$ of all birelational models where the modal accessibility relation is an equivalence relation.\footnote{We treat these two classes of models in a manner analogous to the approach taken in the study of public announcement logic $\PAL$~\cite{vanDitmarsch:et:al:2007}.
In particular, within the context of epistemic reasoning, it is natural to consider the class of all birelational models where the modal accessibility relation is an equivalence relation.}
That is to say, in Section~\ref{section:completeness}, we prove the following completeness results:
\\
\\
{\bf Proposition~$\mathbf{10}$.}
$\L_{\min}$ is complete with respect to ${\mathcal C}_{\min}$.
\\
\\
{\bf Proposition~$\mathbf{12}$.}
$\L_{\partition}$ is complete with respect to ${\mathcal C}_{\partition}$.
%
%
\\
\\
We assume the reader is at home with the basic tools and techniques in intuitionistic logic and modal logic.
See~\cite{Blackburn:et:al:2001,Chagrov:Zakharyaschev:1997,Kracht:1999} for more on them.
See~\cite{Olivetti:2022,IMLA:2017,Stewart:et:al:2018} for an introduction to intuitionistic modal logic and~\cite{vanDitmarsch:et:al:2007,Fagin:et:al:1995,Meyer:VanDerHoeke:1995} for an introduction to epistemic logic and dynamic epistemic logic.
%
%
%
\section{Syntax}\label{section:syntax}
Let $\At$ be a countably infinite set (with typical members called {\em atoms}\/ and denoted $p$, $q$, etc).
Let $\Fo$ be the countably infinite set (with typical members called {\em formulas}\/ and denoted $A$, $B$, etc) defined as follows:
\begin{itemize}
\item $A\ {::=}\ p{\mid}{\bot}{\mid}{\top}{\mid}(A{\vee}A){\mid}(A{\wedge}A){\mid}(A{\rightarrow}A){\mid}(A{\triangleright}A){\mid}{\lbrack A\rbrack}A$,
\end{itemize}
where $p$ ranges over $\At$.
We follow the standard rules for omission of the parentheses.
In particular, for all formulas $A,B,C$, when we write $A{\vee}B{\triangleright}C$, $A{\wedge}B{\triangleright}C$, $A{\triangleright}B{\vee}C$ and $A{\triangleright}B{\wedge}C$, we mean $(A{\vee}B){\triangleright}C$, $(A{\wedge}B){\triangleright}C$, $A{\triangleright}(B{\vee}C)$ and $A{\triangleright}(B{\wedge}C)$.
%
%
%
%
For all formulas $A,B$, we write ${\neg}A$ as the abbreviation of $A{\rightarrow}{\bot}$, $A{\leftrightarrow}B$ as the abbreviation of $(A{\rightarrow}B){\wedge}(B{\rightarrow}A)$, ${\square}A$ as the abbreviation of ${\top}{\triangleright}A$ and ${\blacksquare}A$ as the abbreviation of $A{\triangleright}{\bot}$.
Let $\Sf\ :\ \Fo{\longrightarrow}\wp(\Fo)$ be the function
asociating to each formula $A$ the set $\Sf(A)$ of all $A$'s subformulas.
%
%
%
%
%
%
%
%
%
%
%
%
%
%
%
%
%
%
%
%
%
%
%
%
%
%
A formula $A$ is {\em simple}\/ if for all formulas $B,C$, $B{\triangleright}C{\not\in}\Sf(A)$ and ${\lbrack B\rbrack}C{\not\in}\Sf(A)$.
Let $\Fo_{\simple}$ be the set of all simple formulas.
A {\em simple substitution}\/ is a function $\sigma\ :\ \Fo_{\simple}{\longrightarrow}\Fo$ such that for all simple formulas $A,B$, $\sigma({\bot}){=}{\bot}$, $\sigma({\top}){=}{\top}$, $\sigma(A{\vee}B){=}\sigma(A){\vee}\sigma(B)$, $\sigma(A{\wedge}B){=}\sigma(A){\wedge}\sigma(B)$ and $\sigma(A{\rightarrow}B){=}\sigma(A){\rightarrow}\sigma(B)$.
%
%
%
%
A formula $A$ is {\em direct}\/ if for all formulas $B,C$, if $B{\triangleright}C{\in}
$\linebreak$
\Sf(A)$ then $B{=}{\top}$.
Let $\Fo_{\direct}$ be the set of all direct formulas.
Obviously, direct formulas can also be considered as formulas of the language of $\IPAL$.
A formula $A$ is {\em modal}\/ if for all formulas $B,C$, ${\lbrack B\rbrack}C{\not\in}\Sf(A)$.
Let $\Fo_{\modal}$ be the set of all modal formulas.
Obviously, direct modal formulas can also be considered as $\lozenge$-free formulas of the language of intuitionistic modal logics developed in~\cite{FischerServi:1984,Plotkin:Stirling:1986,Simpson:1994}.
Let $\degre\ :\ \Fo{\longrightarrow}\N$ be the function defined as follows: $\degre(p){=}\degre({\bot}){=}\degre({\top}){=}{0}$, $\degre(A{\vee}B){=}\degre(A{\wedge}
$\linebreak$
B){=}\degre(A{\rightarrow}B){=}\degre(A{\triangleright}B){=}\max\{\degre(A),\degre(B)\}$ and $\degre({\lbrack A\rbrack}B){=}\degre(A){+}{2}
$\linebreak$
{\times}\degre(B){+}{1}$.
Let ${\parallel}{\cdot}{\parallel}\ :\ \Fo{\longrightarrow}\N$ be the function
associating to each formula $A$ the number ${\parallel}A{\parallel}$ of occurrences of symbols in $A$.
%
%
%
%
%
%
%
%
%
%
%
%
%
%
%
%
%
%
%
%
%
%
%
%
Let $\prec$ be the binary relation on $\Fo$ such that for all formulas $A,B$, $A{\prec}B$ if and only if $\degre(A){<}\degre(B)$, or $\degre(A){=}\degre(B)$ and ${\parallel}A{\parallel}{<}{\parallel}B{\parallel}$.
Obviously, $\prec$ is a strict order on $\Fo$.
In other respect, $\Fo$ has no infinite $\prec$-descending chain.
\begin{proposition}\label{proposition:properties:of:prec}
For all formulas $A,B,C$ and for all atoms $p$,
%
%
%
%
$A{\prec}A{\vee}B$, $B{\prec}A{\vee}B$,
%
%
$A{\prec}A{\wedge}B$, $B{\prec}A{\wedge}B$,
%
%
$A{\prec}A{\rightarrow}B$, $B{\prec}A{\rightarrow}B$,
%
%
$A{\prec}A{\triangleright}B$, $B{\prec}A{\triangleright}B$,
%
%
$A{\prec}{\lbrack A\rbrack}p$,
%
%
$A{\prec}{\lbrack A\rbrack}{\bot}$,
%
%
$A{\prec}{\lbrack A\rbrack}(B{\vee}C)$, ${\lbrack A\rbrack}B{\prec}{\lbrack A\rbrack}(B{\vee}C)$, ${\lbrack A\rbrack}C{\prec}{\lbrack A\rbrack}(B{\vee}C)$,
%
%
${\lbrack A\rbrack}B{\prec}{\lbrack A\rbrack}(B{\wedge}C)$, ${\lbrack A\rbrack}C{\prec}{\lbrack A\rbrack}(B
$\linebreak$
{\wedge}C)$,
%
%
${\lbrack A\rbrack}B{\prec}{\lbrack A\rbrack}(B{\rightarrow}C)$, ${\lbrack A\rbrack}C{\prec}{\lbrack A\rbrack}(B{\rightarrow}C)$,
%
%
$A{\prec}{\lbrack A\rbrack}(B{\triangleright}C)$, ${\lbrack A\rbrack}B{\prec}{\lbrack A\rbrack}(B{\triangleright}C)$, ${\lbrack A\rbrack}C
$\linebreak$
{\prec}{\lbrack A\rbrack}(B{\triangleright}C)$ and
%
%
${\lbrack A{\wedge}{\lbrack A\rbrack}B\rbrack}C{\prec}{\lbrack A\rbrack}{\lbrack B\rbrack}C$.
%
%
%
%
\end{proposition}
The intuitive reading of $A{\triangleright}B$ is ``from now on, the agent knows that $B$ holds when $A$ holds''.
As a result, the intuitive reading of ${\square}A$ is ``from now on, the agent knows that $A$ holds'' and the intuitive reading of ${\blacksquare}A$ is ``from now on, the agent knows that $A$ does not hold''.
In other respect, the intuitive reading of ${\lbrack A\rbrack}B$ is ``from now on, $B$ holds after it has been publicly announced that $A$ holds''.
\section{Semantics}\label{section:semantics}
A {\em frame}\/ is a structure of the form $(W,{\leq},{R})$ where $W$ is a nonempty set (with typical members denoted $s$, $t$, etc), $\leq$ is a preorder on $W$ and $R$ is a binary relation on $W$.\footnote{Contrary to what is assumed in many articles, our frames are not necessarily backward and forward confluent, i.e. they are not necessarily such that ${R}{\circ}{\leq}{\subseteq}{\leq}{\circ}{R}$ and ${R^{-1}}{\circ}{\leq}{\subseteq}{\leq}{\circ}{R^{-1}}$.}
Let ${\mathcal C}_{\min}$ be the class of all frames.
Let ${\mathcal C}_{\partition}$ be the class of all frames $(W,{\leq},{R})$ where $R$ is an equivalence relation on $W$, i.e. $R$ is reflexive, symmetric and transitive.
For all frames $(W,{\leq},{R})$, a subset $U$ of $W$ is {\em increasing}\/ if for all $s,t{\in}W$, if $s{\in}U$ and $s{\leq}t$ then $t{\in}U$.
A {\em valuation on a frame $(W,{\leq},{R})$}\/ is a function $V\ :\ \At{\longrightarrow}\wp(W)$ such that for all atoms $p$, $V(p)$ is increasing.
A {\em model}\/ is a structure of the form $(W,{\leq},{R},V)$ where $(W,{\leq},{R})$ is a frame and $V\ :\ \At{\longrightarrow}\wp(W)$ is a valuation on $(W,{\leq},{R})$.
Obviously, for all models $(W,{\leq},{R},V)$ and for all subsets $U$ of $W$, if $U$ is nonempty then the restriction of $(W,{\leq},{R},V)$ to $U$ is a model.
In other respect, for all models $(W,{\leq},{R},V)$ based on a frame in ${\mathcal C}_{\partition}$ and for all subsets $U$ of $W$, if $U$ is nonempty then the restriction of $(W,{\leq},{R},V)$ to $U$ is a model based on a frame in ${\mathcal C}_{\partition}$.
For all formulas $C$, given a model $(W,{\leq},{R},V)$ and $s{\in}W$, the {\em truth condition of $C$ in $(W,{\leq},{R},V)$ at $s$}\/ (in symbols $(W,{\leq},{R},V),s{\models}C$) is defined as follows:\footnote{We omit the truth conditions of $C{=}{\bot}$, $C{=}{\top}$, $C{=}A{\vee}B$ and $C{=}A{\wedge}B$.}
%
%
\begin{itemize}
\item $(W,{\leq},{R},V),s{\models}p$ if and only if $s{\in}V(p)$,
%
%
%
%
%
%
%
%
%
%
\item $(W,{\leq},{R},V),s{\models}A{\rightarrow}B$ if and only if for all $t{\in}W$, if $s{\leq}t$ then $(W,{\leq},{R},V),t{\not\models}A$, or $(W,{\leq},{R},V),t{\models}B$,
\item $(W,{\leq},{R},V),s{\models}A{\triangleright}B$ if and only if for all $t{\in}W$, if $s{\leq}t$ then for all $u{\in}W$, it $t{R}u$ then $(W,{\leq},{R},V),u{\not\models}A$, or $(W,{\leq},{R},V),u{\models}B$,\footnote{This truth condition is reminiscent of the truth condition for strict implication.
See~\cite{Lewis:Langford:1932} for first developments about strict implication and~\cite{deGroot:et:al:2026,Litak:Visser:2018} for recent studies about strict implication.}
\item $(W,{\leq},{R},V),s{\models}{\lbrack A\rbrack}B$ if and only if for all $t{\in}W$, if $s{\leq}t$ then $(W,{\leq},{R},V),t{\not\models}A$, or $(W,{\leq},{R},V)^{+A},t{\models}B$,
\end{itemize}
where $(W,{\leq},{R},V)^{+A}$ denotes the restriction of $(W,{\leq},{R},V)$ to the set of all $v{\in}W$ such that $(W,{\leq},{R},V),v{\models}A$.
The reader may easily verify by induction on the formula $C$ that for all models $(W,{\leq},{R},V)$ and for all $s,t{\in}W$, if $(W,{\leq},{R},V),s{\models}C$ and $s{\leq}t$ then $(W,{\leq},{R},V),t{\models}C$.\footnote{This property of heredity is reminiscent of the property of heredity considered as well in the literature of intuitionistic propositional logic.}\label{footnote:Heredity:Property}
\begin{proposition}\label{proposition:about:truth:sets:of:abbreviations}
Given a model $(W,{\leq},{R},V)$, for all $s{\in}W$ and for all formulas $A,B$, $(W,{\leq},{R},V),s{\models}{\neg}A$ if and only if for all $t{\in}W$, if $s{\leq}t$ then $(W,{\leq},{R},V),t{\not\models}A$, $(W,{\leq},
$\linebreak$
{R},V),s{\models}A{\leftrightarrow}B$ if and only if for all $t{\in}W$, if $s{\leq}t$ then $(W,{\leq},{R},V),t{\models}A$ if and only if $(W,{\leq},{R},V),t{\models}B$, $(W,{\leq},{R},V),s{\models}{\square}A$ if and only if for all $t{\in}W$, if $s{\leq}t$ then for all $u{\in}W$, it $t{R}u$ then $(W,{\leq},{R},V),u{\models}A$ and $(W,{\leq},{R},V),s{\models}{\blacksquare}A$ if and only if for all $t{\in}W$, if $s{\leq}t$ then for all $u{\in}W$, it $t{R}u$ then $(W,{\leq},{R},V),u{\not\models}A$.\footnote{The reader may easily verify that the truth condition for
${\blacksquare}A$ is the one considered in~\cite{Mendler:et:al:2021}.}
\end{proposition}
Obviously, for all formulas $A,B$, for all models $(W,{\leq},{R},V)$ and for all $s{\in}W$, if $s$ belongs to $(W,{\leq},{R},V)^{+A}$ and $(W,{\leq},{R},V)^{+A},s{\models}B$ then $(W,{\leq},{R},V),s{\models}{\lbrack A\rbrack}B$.
Moreover, for all formulas $A,B$, for all models $(W,{\leq},{R},V)$ and for all $s{\in}W$, $s$ belongs to $(W,{\leq},{R},V)^{+A{\wedge}{\lbrack A\rbrack}B}$ if and only if $s$ belongs to $(W,{\leq},{R},V)^{+A}$ and $s$ belongs to $(W,{\leq},{R},V)^{+A+B}$.
In other respect, when these conditions hold, $(W,{\leq},{R},
$\linebreak$
V)^{+A{\wedge}{\lbrack A\rbrack}B}{=}(W,{\leq},{R},V)^{+A+B}$.
A formula $C$ is {\em true in a model $(W,{\leq},{R},V)$}\/ (in symbols $(W,{\leq},{R},V){\models}C$) if for all $s{\in}W$, $s{\models}C$.
A formula $C$ is {\em valid in a frame $(W,{\leq},{R})$}\/ (in symbols $(W,{\leq},{R}){\models}C$) if for all valuations $V\ :\ \At{\longrightarrow}\wp(W)$ on $(W,{\leq},{R})$, $(W,{\leq},{R},V){\models}C$.
A formula $C$ is {\em valid on a class ${\mathcal C}$ of frames}\/ (in symbols ${\mathcal C}{\models}C$) if for all frames $(W,{\leq},{R})$ in ${\mathcal C}$, $(W,{\leq},{R}){\models}C$.
\section{Axiomatization}\label{section:axiomatization}
A {\em logic}\/ is a set $\L$ of formulas such that
%
%
%
%
%
%
%
%
%
%
%
%
\begin{description}
\item[$(\IPL)$] for all simple substitutions $\sigma\ :\ \Fo_{\simple}{\longrightarrow}\Fo$ and for all intuitionistically valid simple formulas $A$, $\L$ contains $\sigma(A)$,
\item[$(\modusponens)$] for all formulas $A,B$, if $A{\in}\L$ and $A{\rightarrow}B{\in}\L$ then $B{\in}\L$,
\item[$(\modalaxiom)$] for all formulas $A,B,C,D$, $\L$ contains
\begin{description}
\item[$(\modalaxiom1)$] $A{\triangleright}A$,
\item[$(\modalaxiom2)$] $(A{\triangleright}C){\wedge}(B{\triangleright}C){\rightarrow}(A{\vee}B{\triangleright}C)$,
\item[$(\modalaxiom3)$] $(A{\triangleright}B){\wedge}(C{\triangleright}D){\rightarrow}(A{\wedge}(B{\rightarrow}C){\triangleright}D)$,
\end{description}
\item[$(\rulenecessitation)$] for all formulas $A,B,C$, if $A{\rightarrow}B{\in}\L$ then $(C{\triangleright}A){\rightarrow}(C{\triangleright}B){\in}\L$ and $(B{\triangleright}C){\rightarrow}
$\linebreak$
(A{\triangleright}C){\in}\L$,
\item[$(\reductionaxiom)$] for all formulas $A,B,C$ and for all atoms $p$, $\L$ contains
\begin{description}
\item[$(\reductionaxiom1)$] ${\lbrack A\rbrack}p{\leftrightarrow}(A{\rightarrow}p)$,
\item[$(\reductionaxiom2)$] ${\lbrack A\rbrack}{\bot}{\leftrightarrow}{\neg}A$,
\item[$(\reductionaxiom3)$] ${\lbrack A\rbrack}{\top}{\leftrightarrow}{\top}$,
\item[$(\reductionaxiom4)$] ${\lbrack A\rbrack}(B{\vee}C){\leftrightarrow}(A{\rightarrow}{\lbrack A\rbrack}B{\vee}{\lbrack A\rbrack}C)$,\footnote{The reader may easily notice the slight difference between $(\reductionaxiom4)$ and the reduction axiom for ``disjunction'' ${\lbrack A\rbrack}(B{\vee}C){\leftrightarrow}{\lbrack A\rbrack}B{\vee}{\lbrack A\rbrack}C$ considered in the literature of $\IPAL$~\cite{Ma:Palmigiano:Sadrzadeh:2014,Maffezioli:Negri:2011}.
Indeed, the reduction axiom ${\lbrack A\rbrack}(B{\vee}C){\leftrightarrow}{\lbrack A\rbrack}B{\vee}{\lbrack A\rbrack}C$ does not respect the constructive character of ``disjunction'' seeing that the following situation can be realized: $\mathbf{(i)}$~in all future situations where $A$ holds, $B$ holds or $C$ holds after the announcement of $A$ while $\mathbf{(ii)}$~it cannot be determined, before $A$ holds, which formula among $B$ and $C$ will hold after the announcement of $A$.}
\item[$(\reductionaxiom5)$] ${\lbrack A\rbrack}(B{\wedge}C){\leftrightarrow}{\lbrack A\rbrack}B{\wedge}{\lbrack A\rbrack}C$,
\item[$(\reductionaxiom6)$] ${\lbrack A\rbrack}(B{\rightarrow}C){\leftrightarrow}({\lbrack A\rbrack}B{\rightarrow}{\lbrack A\rbrack}C)$,
\item[$(\reductionaxiom7)$] ${\lbrack A\rbrack}(B{\triangleright}C){\leftrightarrow}(A{\rightarrow}(A{\wedge}{\lbrack A\rbrack}B{\triangleright}{\lbrack A\rbrack}C))$,\footnote{When $B{=}{\top}$, the reader may easily notice the slight difference between ${\lbrack A\rbrack}({\top}{\triangleright}C){\leftrightarrow}(A{\rightarrow}(A{\wedge}{\lbrack A\rbrack}{\top}{\triangleright}{\lbrack A\rbrack}C))$~---~semantically equivalent with ${\lbrack A\rbrack}{\square}C{\leftrightarrow}(A{\rightarrow}(A{\triangleright}{\lbrack A\rbrack}C))$~---~and the reduction axiom for ``necessity'' ${\lbrack A\rbrack}{\square}C{\leftrightarrow}(A{\rightarrow}{\square}{\lbrack A\rbrack}C)$~---~syntactically equal to ${\lbrack A\rbrack}{\square}C{\leftrightarrow}(A{\rightarrow}({\top}{\triangleright}{\lbrack A\rbrack}C))$~---~considered in the literature of $\IPAL$~\cite{Ma:Palmigiano:Sadrzadeh:2014,Maffezioli:Negri:2011}.
Indeed, the reduction axiom ${\lbrack A\rbrack}{\square}C{\leftrightarrow}(A{\rightarrow}{\square}{\lbrack A\rbrack}C)$ does not respect the ability of the agent to conceive alternative historical developments seeing that the following situation can be realized: $\mathbf{(i)}$~in all future situations where $A$ holds, after the announcement of $A$, the agent knows that $C$ holds while $\mathbf{(ii)}$~in some future situation where $A$ holds, the agent does not know that in all future situations where $A$ holds, $C$ holds after the announcement of $A$.}
\item[$(\reductionaxiom8)$] ${\lbrack A\rbrack}{\lbrack B\rbrack}C{\leftrightarrow}{\lbrack A{\wedge}{\lbrack A\rbrack}B\rbrack}C$.
\end{description}
\end{description}
Let $\Sigma_{\reflexive}$ be the least set of formulas containing all formulas of the form
\begin{description}
\item[$(\modalaxiom4)$] $(A{\triangleright}B){\rightarrow}(A{\rightarrow}B)$,
\end{description}
where $A,B$ range over $\Fo$.
Let $\Sigma_{\symmetric}$ be the least set of formulas containing all formulas of the form
\begin{description}
\item[$(\modalaxiom5)$] $(A{\triangleright}B{\vee}(C{\triangleright}D)){\rightarrow}((D{\rightarrow}(A{\triangleright}B)){\rightarrow}(C{\rightarrow}(A{\triangleright}B)))$,
\end{description}
where $A,B,C,D$ range over $\Fo$.
%
%
Let $\L_{\min}$ be the least logic.
Let $\L_{\partition}$ be the least logic containing $\Sigma_{\reflexive}{\cup}\Sigma_{\symmetric}$.
A logic $\L$ is {\em consistent}\/ if ${\bot}{\not\in}\L$.
Obviously, for all logics $\L$, $\L$ is consistent if and only if $\L{\not=}\Fo$.
A logic $\L$ is {\em sound with respect to a class ${\mathcal C}$ of frames}\/ if for all formulas $A$, if $A{\in}\L$ then ${\mathcal C}{\models}A$.
A logic $\L$ is {\em complete with respect to a class ${\mathcal C}$ of frames}\/ if for all formulas $A$, if ${\mathcal C}{\models}A$ then $A{\in}\L$.
%
%
%
%
%
%
\\
\\
Proposition~\ref{proposition:logic:replacement:by:equivalent:formulas} says that the standard rule of replacement by equivalent is admissible in any logic.
\begin{proposition}\label{proposition:logic:replacement:by:equivalent:formulas}
Let $\L$ be a logic.
For all formulas $A,B,C,D$, if $A{\leftrightarrow}B{\in}\L$ and $C{\leftrightarrow}D
$\linebreak$
{\in}\L$ then $(A{\vee}C){\leftrightarrow}(B{\vee}D){\in}\L$, $(A{\wedge}C){\leftrightarrow}(B{\wedge}D){\in}\L$, $(A{\rightarrow}C){\leftrightarrow}(B{\rightarrow}D){\in}\L$ and $(A{\triangleright}
$\linebreak$
C){\leftrightarrow}(B{\triangleright}D){\in}\L$.
\end{proposition}
\begin{proof}
Let $A,B,C,D$ be formulas.
Suppose $A{\leftrightarrow}B{\in}\L$ and $C{\leftrightarrow}D{\in}\L$.
We only consider the last fact.
Since $A{\leftrightarrow}B{\in}\L$ and $C{\leftrightarrow}D{\in}\L$, therefore $A{\rightarrow}B{\in}\L$, $B{\rightarrow}A{\in}\L$, $C{\rightarrow}D
$\linebreak$
{\in}\L$ and $D{\rightarrow}C{\in}\L$.
Hence, using $(\rulenecessitation)$, $(A{\triangleright}C){\rightarrow}(B{\triangleright}C){\in}\L$, $(B{\triangleright}C){\rightarrow}(B{\triangleright}D){\in}\L$, $(B{\triangleright}D){\rightarrow}(A{\triangleright}D){\in}\L$ and $(A{\triangleright}D){\rightarrow}(A{\triangleright}C){\in}\L$.
Thus, $(A{\triangleright}C){\rightarrow}(B{\triangleright}D){\in}\L$ and $(B{\triangleright}D)
$\linebreak$
{\rightarrow}(A{\triangleright}C){\in}\L$.
Consequently, $(A{\triangleright}C){\leftrightarrow}(B{\triangleright}D){\in}\L$.
\medskip
\end{proof}
%
%
%
%
Proposition~\ref{proposition:logic:reduction:is:always:possible} says that as for public announcement logic $\PAL$~\cite{vanDitmarsch:et:al:2007}, reduction axioms provide in any logic, a deterministic algorithm producing given a formula, a provably equivalent modal formula.
\begin{proposition}\label{proposition:logic:reduction:is:always:possible}
Let $\L$ be a logic.
There exists a deterministic algorithm producing given a formula $A$, a modal formula $B$ such that $A{\leftrightarrow}B{\in}\L$.
\end{proposition}
\begin{proof}
By Proposition~\ref{proposition:properties:of:prec}, let $\tau\ :\ \Fo{\longrightarrow}\Fo_{\modal}$ be the function defined as follows: $\tau(p){=}p$, $\tau({\bot}){=}{\bot}$, $\tau({\top}){=}{\top}$, $\tau(A{\vee}B){=}\tau(A){\vee}\tau(B)$, $\tau(A{\wedge}B){=}\tau(A){\wedge}\tau(B)$, $\tau(A{\rightarrow}
$\linebreak$
B){=}\tau(A){\rightarrow}\tau(B)$, $\tau(A{\triangleright}B){=}\tau(A){\triangleright}\tau(B)$, $\tau({\lbrack A\rbrack}p){=}\tau(A){\rightarrow}p$, $\tau({\lbrack A\rbrack}{\bot}){=}{\neg}\tau(A)$,
\linebreak$
\tau({\lbrack A\rbrack}{\top}){=}{\top}$, $\tau({\lbrack A\rbrack}(B{\vee}C)){=}\tau(A){\rightarrow}\tau({\lbrack A\rbrack}B){\vee}\tau({\lbrack A\rbrack}C)$, $\tau({\lbrack A\rbrack}(B{\wedge}C)){=}\tau({\lbrack A\rbrack}B){\wedge}
$\linebreak$
\tau({\lbrack A\rbrack}C)$, $\tau({\lbrack A\rbrack}(B{\rightarrow}C)){=}\tau({\lbrack A\rbrack}B){\rightarrow}\tau({\lbrack A\rbrack}C)$, $\tau({\lbrack A\rbrack}(B{\triangleright}C)){=}\tau(A){\rightarrow}(\tau(A){\wedge}
$\linebreak$
\tau({\lbrack A\rbrack}B){\triangleright}\tau({\lbrack A\rbrack}C))$ and $\tau({\lbrack A\rbrack}{\lbrack B\rbrack}C){=}\tau({\lbrack A{\wedge}{\lbrack A\rbrack}B\rbrack}C)$.
The reader may easily verify by induction on the formula $A$ that $A{\leftrightarrow}\tau(A){\in}\L$.
\medskip
\end{proof}
Our goal is to show the equivalences ``for all formulas $A$, $A{\in}\L_{\min}$ if and only if ${\mathcal C}_{\min}{\models}A$'' and ``for all formulas $A$, $A{\in}\L_{\partition}$ if and only if ${\mathcal C}_{\partition}{\models}A$''.
With Propositions~\ref{proposition:soundness:of:L:min} and~\ref{proposition:soundness:of:L:partition}, we show the implications ``for all formulas $A$, if $A{\in}\L_{\min}$ then ${\mathcal C}_{\min}{\models}A$'' and ``for all formulas $A$, if $A{\in}\L_{\partition}$ then ${\mathcal C}_{\partition}{\models}A$''.
The converse implications ``for all formulas $A$, if ${\mathcal C}_{\min}{\models}A$ then $A{\in}\L_{\min}$'' and ``for all formulas $A$, if ${\mathcal C}_{\partition}{\models}A$ then $A{\in}\L_{\partition}$'' will be shown in Section~\ref{section:completeness} with Propositions~\ref{proposition:completeness:of:L:min} and~\ref{proposition:completeness:of:L:partition}.
\begin{proposition}[Soundness of $\L_{\min}$]\label{proposition:soundness:of:L:min}
$\L_{\min}$ is sound with respect to ${\mathcal C}_{\min}$.
%
%
\end{proposition}
\begin{proof}
It suffices to demonstrate that $(\IPL)$, $(\modusponens)$, $(\modalaxiom1)$, $(\modalaxiom2)$, $(\modalaxiom3)$,
\linebreak$
(\rulenecessitation)$ and $(\reductionaxiom1)$--$(\reductionaxiom8)$ give rise to ${\mathcal C}_{\min}$-valid formulas.
We only consider the ``left to right'' direction of the cases of $(\reductionaxiom7)$ and $(\reductionaxiom8)$.
\\
{\bf Case of $(\reductionaxiom7)$:}
For the sake of the contradiction, suppose there exists formulas $A,B,
$\linebreak$
C$ such that ${\lbrack A\rbrack}(B{\triangleright}C){\rightarrow}(A{\rightarrow}(A{\wedge}{\lbrack A\rbrack}B{\triangleright}{\lbrack A\rbrack}C))$ is not ${\mathcal C}_{\min}$-valid.
Hence, there exists a model $(W,{\leq},{R},V)$ and there exists $s{\in}W$ such that $(W,{\leq},{R},V),s{\models}{\lbrack A\rbrack}(B{\triangleright}C)$, $(W,{\leq},{R},V),s{\models}A$ and $(W,{\leq},{R},V),s{\not\models}A{\wedge}{\lbrack A\rbrack}B{\triangleright}{\lbrack A\rbrack}C$.
Thus, $(W,{\leq},{R},V)^{+A},s{\models}
$\linebreak$
B{\triangleright}C$.
Moreover, there exists $t{\in}W$ such that $s{\leq}t$ and there exists $u{\in}W$ such that $t{R}u$, $(W,{\leq},{R},V),u{\models}A{\wedge}{\lbrack A\rbrack}B$ and $(W,{\leq},{R},V),u{\not\models}{\lbrack A\rbrack}C$.
Since $(W,{\leq},{R},V),s{\models}A$,
\linebreak
therefore $(W,{\leq},{R},V),t{\models}A$.\footnote{Here, we use the property of heredity mentioned in Footnote~$4$.
From now on, we will use the property of heredity without explicitly mentioning it.}
Moreover, $(W,{\leq},{R},V),u{\models}A$ and $(W,{\leq},{R},V),u{\models}
$\linebreak$
{\lbrack A\rbrack}B$.
Consequently, $(W,{\leq},{R},V)^{+A},u{\models}B$.
Since $(W,{\leq},{R},V)^{+A},s{\models}B{\triangleright}C$, $s{\leq}t$, $t{R}u$, $(W,{\leq},{R},V),t{\models}A$ and $(W,{\leq},{R},V),u{\models}A$, therefore $(W,{\leq},{R},V)^{+A},u{\models}C$.
Hence, $(W,{\leq},{R},V),u{\models}{\lbrack A\rbrack}C$: a contradiction.
\\
{\bf Case of $(\reductionaxiom8)$:}
For the sake of the contradiction, suppose there exists formulas $A,B,
$\linebreak$
C$ such that ${\lbrack A\rbrack}{\lbrack B\rbrack}C{\rightarrow}{\lbrack A{\wedge}{\lbrack A\rbrack}B\rbrack}C$ is not ${\mathcal C}_{\min}$-valid.
Thus, there exists a model $(W,{\leq},
$\linebreak$
{R},V)$ and there exists $s{\in}W$ such that $(W,{\leq},{R},V),s{\models}{\lbrack A\rbrack}{\lbrack B\rbrack}C$ and $(W,{\leq},{R},V),s{\not\models}
$\linebreak$
{\lbrack A{\wedge}{\lbrack A\rbrack}B\rbrack}C$.
Consequently, there exists $t{\in}W$ such that $s{\leq}t$, $(W,{\leq},{R},V),t{\models}A{\wedge}{\lbrack A\rbrack}B$ and $(W,{\leq},{R},V)^{+A{\wedge}{\lbrack A\rbrack}B},t{\not\models}C$.
Hence, $(W,{\leq},{R},V),t{\models}A$ and $(W,{\leq},{R},V),t{\models}
$\linebreak$
{\lbrack A\rbrack}B$.
Thus, $(W,{\leq},{R},V)^{+A},t{\models}B$.
Moreover, since $(W,{\leq},{R},V),s{\models}{\lbrack A\rbrack}{\lbrack B\rbrack}C$ and $s{\leq}t$, therefore $(W,{\leq},{R},V)^{+A},t{\models}{\lbrack B\rbrack}C$.
Consequently, $(W,{\leq},{R},V)^{+A+B},t{\models}C$.
\linebreak
Since $t$ belongs to $(W,{\leq},{R},V)^{+A{\wedge}{\lbrack A\rbrack}B}$, $t$ belongs to $(W,{\leq},{R},V)^{+A}$ and $t$ belongs to $(W,{\leq},{R},V)^{+A+B}$, therefore $(W,{\leq},{R},V)^{+A{\wedge}{\lbrack A\rbrack}B}{=}(W,{\leq},{R},V)^{+A+B}$.
Since
\linebreak$
(W,{\leq},{R},V)^{+A+B},t{\models}C$, therefore $(W,{\leq},{R},V)^{+A{\wedge}{\lbrack A\rbrack}B},t{\models}C$: a contradiction.
\medskip
\end{proof}
\begin{proposition}[Soundness of $\L_{\partition}$]\label{proposition:soundness:of:L:partition}
$\L_{\partition}$ is sound with respect to ${\mathcal C}_{\partition}$.
%
%
\end{proposition}
\begin{proof}
It suffices to demonstrate that $(\IPL)$, $(\modusponens)$, $(\modalaxiom1)$--$(\modalaxiom5)$, $(\rulenecessitation)$ and $(\reductionaxiom1)$--$(\reductionaxiom8)$ give rise to ${\mathcal C}_{\partition}$-valid formulas.
We only consider the cases of $(\modalaxiom4)$ and $(\modalaxiom5)$.
\\
{\bf Case of $(\modalaxiom4)$:}
For the sake of the contradiction, suppose there exists formulas $A,B$ such that $(A{\triangleright}B){\rightarrow}(A{\rightarrow}B)$ is not ${\mathcal C}_{\partition}$-valid.
Hence, there exists a model $(W,{\leq},{R},V)$ based on a frame in ${\mathcal C}_{\partition}$ and there exists $s{\in}W$ such that $(W,{\leq},{R},V),s{\models}A{\triangleright}B$, $(W,{\leq},{R},V),s{\models}A$ and $(W,{\leq},{R},V),s{\not\models}B$.
Since $R$ is reflexive, therefore $(W,{\leq},{R},
$\linebreak$
V),s{\models}B$: a contradiction.
\\
{\bf Case of $(\modalaxiom5)$:}
For the sake of the contradiction, suppose there exists formulas $A,B,C,D$ such that $(A{\triangleright}B{\vee}(C{\triangleright}D)){\rightarrow}((D{\rightarrow}(A{\triangleright}B)){\rightarrow}(C{\rightarrow}(A{\triangleright}B)))$ is not
\linebreak$
{\mathcal C}_{\partition}$-valid.
Thus, there exists a model $(W,{\leq},{R},V)$ based on a frame in ${\mathcal C}_{\partition}$ and there exists $s{\in}W$ such that $(W,{\leq},{R},V),s{\models}A{\triangleright}B{\vee}(C{\triangleright}D)$, $(W,{\leq},{R},V),s{\models}D{\rightarrow}(A{\triangleright}B)$, $(W,{\leq},{R},V),s{\models}C$ and $(W,{\leq},{R},V),s{\not\models}A{\triangleright}B$.
Consequently, there exists $t{\in}W$ such that $s{\leq}t$ and there exists $u{\in}W$ such that $t{R}u$, $(W,{\leq},{R},V),u{\models}A$ and $(W,{\leq},{R},V),u
$\linebreak$
{\not\models}B$.
Since $(W,{\leq},{R},V),s{\models}C$, therefore $(W,{\leq},{R},V),t{\models}C$.
Moreover, since $(W,{\leq},
$\linebreak$
{R},V),s{\models}A{\triangleright}B{\vee}(C{\triangleright}D)$, therefore $(W,{\leq},{R},V),u{\models}B{\vee}(C{\triangleright}D)$.
Thus, $(W,{\leq},{R},V),
$\linebreak$
u{\models}B$, or $(W,{\leq},{R},V),u{\models}C{\triangleright}D$.
Since $(W,{\leq},{R},V),u{\not\models}B$, therefore $(W,{\leq},{R},V),
$\linebreak$
u{\models}C{\triangleright}D$.
Since $R$ is symmetric, $t{R}u$ and $(W,{\leq},{R},V),t{\models}C$, therefore $(W,{\leq},{R},V),
$\linebreak$
t{\models}D$.
Since $(W,{\leq},{R},V),s{\models}D{\rightarrow}(A{\triangleright}B)$ and $s{\leq}t$, therefore $(W,{\leq},{R},V),t{\models}(A{\triangleright}B)$.
Since $t{R}u$ and $(W,{\leq},{R},V),u{\models}A$, therefore $(W,{\leq},{R},V),u{\models}B$: a contradiction.
\medskip
\end{proof}
\section{Theories}\label{section:theories}
In this section, we provide definitions and results concerning the elements composing the canonical frame of $\L_{\min}$ and the canonical frame of $\L_{\partition}$ defined in Section~\ref{section:completeness}.
\\
\\
Let $\L$ be a consistent logic.
A {\em $\L$-theory}\/ is a set $\Gamma$ of formulas such that $\L{\subseteq}\Gamma$ and for all formulas $A,B$, if $A{\in}\Gamma$ and $A{\rightarrow}B{\in}\Gamma$ then $B{\in}\Gamma$.
%
%
%
%
A $\L$-theory $\Gamma$ is {\em proper}\/ if ${\bot}{\not\in}\Gamma$.
Obviously, for all $\L$-theories $\Gamma$, $\Gamma$ is proper if and only if $\Gamma{\not=}\Fo$.
A proper $\L$-theory $\Gamma$ is {\em prime}\/ if for all formulas $A,B$, if $A{\vee}B{\in}\Gamma$ then $A{\in}\Gamma$, or $B{\in}\Gamma$.
For all $\L$-theories $\Gamma$ and for all formulas $A$, let $\Gamma{+}A{=}\{B{\in}\Fo$: $A{\rightarrow}B{\in}\Gamma\}$.
It immediately follows from these definitions that for all $\L$-theories $\Gamma$ and for all formulas $A$, $\Gamma{+}A$ is the least $\L$-theory containing $\Gamma$ and $A$.
%
%
%
%
%
%
%
%
%
%
%
%
%
%
Imitating the standard results about theories in intuitionistic propositional logic, the reader may easily demonstrate that for all $\L$-theories $\Gamma$ and for all formulas $A$, if $A{\not\in}\Gamma$ then there exists a prime $\L$-theory $\Delta$ such that $\Gamma{\subseteq}\Delta$ and $A{\not\in}\Delta$ and for all $\L$-theories $\Gamma$ and for all formulas $A,B$, if $A{\rightarrow}B{\not\in}\Gamma$ then there exists a prime $\L$-theory $\Delta$ such that $\Gamma{\subseteq}\Delta$, $A{\in}\Delta$ and $B{\not\in}\Delta$.
%
%
%
%
\\
\\
In order to show the completeness of $\L_{\min}$ and the completeness of $\L_{\partition}$ in Section~\ref{section:completeness}, we show with Propositions~\ref{proposition:triangleright:there:exists:a:prime:theory:containing} and~\ref{proposition:triangleright:there:exists:a:prime:theory:containing:symmetric:case} results that will play for $\L_{\min}$ and for $\L_{\partition}$ the role played by the Existence Lemma in classical modal logic~\cite[Lemma~$4.20$]{Blackburn:et:al:2001}.
\begin{proposition}\label{proposition:triangleright:there:exists:a:prime:theory:containing}
For all $\L$-theories $\Gamma$ and for all formulas $A,B$, if $A{\triangleright}B{\not\in}\Gamma$ then there exists a prime $\L$-theory $\Delta$ such that $\mathbf{(i)}$~for all formulas $C,D$, if $C{\triangleright}D{\in}\Gamma$ then $C{\not\in}\Delta$, or $D{\in}\Delta$, $\mathbf{(ii)}$~$A{\in}\Delta$ and $\mathbf{(iii)}$~$B{\not\in}\Delta$.
%
%
\end{proposition}
\begin{proof}\footnote{The reader may easily see how $(\modalaxiom1)$, $(\modalaxiom2)$, $(\modalaxiom3)$ and $(\rulenecessitation)$ are used in this proof.}
Let $\Gamma$ be a $\L$-theory and $A,B$ be formulas.
Suppose $A{\triangleright}B{\not\in}\Gamma$.
Notice that using $(\modalaxiom1)$, $B{\triangleright}B{\in}\Gamma$.
Let ${\mathcal S}{=}\{(\Delta,\Lambda)$: $\Delta$ is a $\L$-theory and $\Lambda$ is a set of formulas such that for all $C{\in}\Delta$, for all $n{\in}\N$ and for all $D_{1},\ldots,D_{n}{\in}\Lambda$, $C{\triangleright}(D_{1}{\vee}\ldots(D_{n}{\vee}{\bot})\ldots){\not\in}\Gamma$, $A{\in}\Delta$ and $B{\in}\Lambda\}$.
\\
{\bf We claim that $(\L{+}A,\{B\}){\in}{\mathcal S}$.}
If not, there exists $C{\in}\L{+}A$, there exists $n{\in}\N$ and there exists $D_{1},\ldots,D_{n}{\in}\{B\}$ such that $C{\triangleright}(D_{1}{\vee}\ldots(D_{n}{\vee}{\bot})\ldots){\in}\Gamma$.
Hence, $A{\rightarrow}C{\in}
$\linebreak$
\L$.
Moreover, $(D_{1}{\vee}\ldots(D_{n}{\vee}{\bot})\ldots){\rightarrow}B{\in}\L$.
Thus, using $(\rulenecessitation)$, $(C{\triangleright}(D_{1}{\vee}\ldots(D_{n}{\vee}
$\linebreak$
{\bot})\ldots)){\rightarrow}(A{\triangleright}(D_{1}{\vee}\ldots(D_{n}{\vee}{\bot})\ldots)){\in}\L$ and $(A{\triangleright}(D_{1}{\vee}\ldots(D_{n}{\vee}{\bot})\ldots)){\rightarrow}(A{\triangleright}B){\in}
$\linebreak$
\L$.
Consequently, $(C{\triangleright}(D_{1}{\vee}\ldots(D_{n}{\vee}{\bot})\ldots)){\rightarrow}(A{\triangleright}B){\in}\L$.
Since $C{\triangleright}(D_{1}{\vee}\ldots(D_{n}{\vee}
$\linebreak$
{\bot})\ldots){\in}\Gamma$, therefore $A{\triangleright}B{\in}\Gamma$: a contradiction.
Therefore $(\L{+}A,\{B\}){\in}{\mathcal S}$.
Hence, ${\mathcal S}$ is nonempty.
Moreover, obviously, for all nonempty chains $(\Delta_{i},\Lambda_{i})_{i{\in}I}$ of elements of ${\mathcal S}$, $(\bigcup\{\Delta_{i}$: $i{\in}I\},\bigcup\{\Lambda_{i}$: $i{\in}I\})$ is an element of ${\mathcal S}$.
Thus, by Kuratowski-Zorn Lemma, ${\mathcal S}$ possesses a maximal element $(\Delta,\Lambda)$.
Consequently, $\Delta$ is a $\L$-theory and $\Lambda$ is a set of formulas such that
\begin{description}
\item[$({\ast})$] for all $C{\in}\Delta$, for all $n{\in}\N$ and for all $D_{1},\ldots,D_{n}{\in}\Lambda$, $C{\triangleright}(D_{1}{\vee}\ldots(D_{n}{\vee}{\bot})\ldots){\not\in}
$\linebreak$
\Gamma$,
\item[$({\ast\ast})$] $A{\in}\Delta$,
\item[$({\ast\ast\ast})$] $B{\in}\Lambda$.
\end{description}
{\bf We claim that $\Delta$ is proper.}
If not, $\Delta{=}\Fo$.
Hence, $B{\in}\Delta$.
Since~$({\ast})$ and~$({\ast}{\ast}{\ast})$, therefore $B{\triangleright}B{\not\in}\Gamma$: a contradiction.
Therefore $\Delta$ is proper.
\\
{\bf We claim that $\Delta$ is prime.}
If not, there exists formulas $E,F$ such that $E{\vee}F{\in}\Delta$, $E{\not\in}\Delta$ and $F{\not\in}\Delta$.
Thus, by the maximality of $(\Delta,\Lambda)$ in ${\mathcal S}$, $(\Delta{+}E,\Lambda){\not\in}{\mathcal S}$ and $(\Delta{+}F,\Lambda){\not\in}{\mathcal S}$.
Consequently, there exists $C^{\prime}{\in}\Delta{+}E$, there exists $n^{\prime}{\in}\N$ and there exists $D^{\prime}_{1},\ldots,D^{\prime}_{n^{\prime}}{\in}
$\linebreak$
\Lambda$ such that $C^{\prime}{\triangleright}(D^{\prime}_{1}{\vee}\ldots(D^{\prime}_{n^{\prime}}{\vee}{\bot})\ldots){\in}\Gamma$ and there exists $C^{\prime\prime}{\in}\Delta{+}F$, there exists $n^{\prime\prime}{\in}\N$ and there exists $D^{\prime\prime}_{1},\ldots,D^{\prime\prime}_{n^{\prime\prime}}{\in}\Lambda$ such that $C^{\prime\prime}{\triangleright}(D^{\prime\prime}_{1}{\vee}\ldots(D^{\prime\prime}_{n^{\prime\prime}}{\vee}{\bot})\ldots){\in}\Gamma$.
\linebreak
Hence, $E{\rightarrow}C^{\prime}{\in}\Delta$ and $F{\rightarrow}C^{\prime\prime}{\in}\Delta$.
Thus, $E{\vee}F{\rightarrow}C^{\prime}{\vee}C^{\prime\prime}{\in}\Delta$.
Since $E{\vee}F{\in}\Delta$, therefore $C^{\prime}{\vee}C^{\prime\prime}{\in}\Delta$.
Since $(D^{\prime}_{1}{\vee}\ldots(D^{\prime}_{n^{\prime}}{\vee}{\bot})\ldots){\rightarrow}(D^{\prime}_{1}{\vee}\ldots(D^{\prime}_{n^{\prime}}{\vee}(D^{\prime\prime}_{1}{\vee}\ldots(D^{\prime\prime}_{n^{\prime\prime}}{\vee}
$\linebreak$
{\bot})\ldots))\ldots){\in}\L$ and $(D^{\prime\prime}_{1}{\vee}\ldots(D^{\prime\prime}_{n^{\prime\prime}}{\vee}{\bot})\ldots){\rightarrow}(D^{\prime}_{1}{\vee}\ldots(D^{\prime}_{n^{\prime}}{\vee}(D^{\prime\prime}_{1}{\vee}\ldots(D^{\prime\prime}_{n^{\prime\prime}}{\vee}
$\linebreak$
{\bot})\ldots))\ldots){\in}\L$, therefore using $(\rulenecessitation)$, $(C^{\prime}{\triangleright}(D^{\prime}_{1}{\vee}\ldots(D^{\prime}_{n^{\prime}}{\vee}{\bot})\ldots)){\rightarrow}(C^{\prime}{\triangleright}(D^{\prime}_{1}{\vee}
$\linebreak$
\ldots(D^{\prime}_{n^{\prime}}{\vee}(D^{\prime\prime}_{1}{\vee}\ldots(D^{\prime\prime}_{n^{\prime\prime}}{\vee}{\bot})\ldots))\ldots)){\in}\L$ and $(C^{\prime\prime}{\triangleright}(D^{\prime\prime}_{1}{\vee}\ldots(D^{\prime\prime}_{n^{\prime\prime}}{\vee}{\bot})\ldots)){\rightarrow}(C^{\prime\prime}
$\linebreak$
{\triangleright}(D^{\prime}_{1}{\vee}\ldots(D^{\prime}_{n^{\prime}}{\vee}(D^{\prime\prime}_{1}{\vee}\ldots(D^{\prime\prime}_{n^{\prime\prime}}{\vee}{\bot})\ldots))\ldots)){\in}\L$.
Since $C^{\prime}{\triangleright}(D^{\prime}_{1}{\vee}\ldots(D^{\prime}_{n^{\prime}}{\vee}{\bot})\ldots)
$\linebreak$
{\in}\Gamma$ and $C^{\prime\prime}{\triangleright}(D^{\prime\prime}_{1}{\vee}\ldots(D^{\prime\prime}_{n^{\prime\prime}}{\vee}{\bot})\ldots){\in}\Gamma$, therefore $C^{\prime}{\triangleright}(D^{\prime}_{1}{\vee}\ldots(D^{\prime}_{n^{\prime}}{\vee}(D^{\prime\prime}_{1}{\vee}\ldots(D^{\prime\prime}_{n^{\prime\prime}}
$\linebreak$
{\vee}{\bot})\ldots))\ldots){\in}\Gamma$ and $C^{\prime\prime}{\triangleright}(D^{\prime}_{1}{\vee}\ldots(D^{\prime}_{n^{\prime}}{\vee}(D^{\prime\prime}_{1}{\vee}\ldots(D^{\prime\prime}_{n^{\prime\prime}}{\vee}{\bot})\ldots))\ldots){\in}\Gamma$.
\linebreak
Consequently, using $(\modalaxiom2)$, $C^{\prime}{\vee}C^{\prime\prime}{\triangleright}(D^{\prime}_{1}{\vee}\ldots(D^{\prime}_{n^{\prime}}{\vee}(D^{\prime\prime}_{1}{\vee}\ldots(D^{\prime\prime}_{n^{\prime\prime}}{\vee}{\bot})\ldots))\ldots){\in}
$\linebreak$
\Gamma$.
Since~$({\ast})$, $D^{\prime}_{1},\ldots,D^{\prime}_{n^{\prime}}{\in}\Lambda$, $D^{\prime\prime}_{1},\ldots,D^{\prime\prime}_{n^{\prime\prime}}{\in}\Lambda$ and $C^{\prime}{\vee}C^{\prime\prime}{\in}\Delta$, therefore $C^{\prime}{\vee}C^{\prime\prime}{\triangleright}
$\linebreak$
(D^{\prime}_{1}{\vee}\ldots(D^{\prime}_{n^{\prime}}{\vee}(D^{\prime\prime}_{1}{\vee}\ldots(D^{\prime\prime}_{n^{\prime\prime}}{\vee}{\bot})\ldots))\ldots){\not\in}\Gamma$: a contradiction.
\\
{\bf We claim that for all formulas $C,D$, if $C{\triangleright}D{\in}\Gamma$ then $C{\not\in}\Delta$, or $D{\in}\Delta$.}
If not, there exists formulas $C,D$ such that $C{\triangleright}D{\in}\Gamma$, $C{\in}\Delta$ and $D{\not\in}\Delta$.
Hence, by the maximality of $(\Delta,\Lambda)$ in ${\mathcal S}$, $(\Delta{+}D,\Lambda){\not\in}{\mathcal S}$.
Thus, there exists $C^{\prime}{\in}\Delta{+}D$, there exists $n^{\prime}{\in}\N$ and there exists $D^{\prime}_{1},\ldots,D^{\prime}_{n^{\prime}}{\in}\Lambda$ such that $C^{\prime}{\triangleright}(D^{\prime}_{1}{\vee}\ldots(D^{\prime}_{n^{\prime}}{\vee}{\bot})\ldots){\in}\Gamma$.
Consequently, $D{\rightarrow}C^{\prime}{\in}\Delta$.
Since $C{\in}\Delta$, therefore $C{\wedge}(D{\rightarrow}C^{\prime}){\in}\Delta$.
Since $({\ast})$ and $D^{\prime}_{1},\ldots,D^{\prime}_{n^{\prime}}{\in}\Lambda$, therefore $C{\wedge}(D{\rightarrow}C^{\prime}){\triangleright}(D^{\prime}_{1}{\vee}\ldots(D^{\prime}_{n^{\prime}}{\vee}{\bot})\ldots){\not\in}\Gamma$.
Since $C{\triangleright}D{\in}\Gamma$ and $C^{\prime}{\triangleright}(D^{\prime}_{1}
$\linebreak$
{\vee}\ldots(D^{\prime}_{n^{\prime}}{\vee}{\bot})\ldots){\in}\Gamma$, therefore using $(\modalaxiom3)$, $C{\wedge}(D{\rightarrow}C^{\prime}){\triangleright}(D^{\prime}_{1}{\vee}\ldots(D^{\prime}_{n^{\prime}}{\vee}{\bot})\ldots)
$\linebreak$
{\in}\Gamma$: a contradiction.
\\
{\bf We claim that $B{\not\in}\Delta$.}
If not, since~$({\ast})$ and~$({\ast}{\ast}{\ast})$, therefore $B{\triangleright}B{\not\in}\Gamma$: a contradiction.
Therefore $B{\not\in}\Delta$.
\medskip
\end{proof}
\begin{proposition}\label{proposition:triangleright:there:exists:a:prime:theory:containing:symmetric:case}
If $\L$ contains $\L_{\partition}$ then for all $\L$-theories $\Gamma$ and for all formulas $A,B$, if $A{\triangleright}B{\not\in}\Gamma$ then there exists a prime $\L$-theory $\Delta$ such that $\Gamma{\subseteq}\Delta$ and there exists a prime $\L$-theory $\Lambda$ such that $\mathbf{(i)}$~for all formulas $C,D$, if $C{\triangleright}D{\in}\Delta$ then $C{\not\in}\Lambda$, or $D{\in}\Lambda$, $\mathbf{(ii)}$~for all formulas $C,D$, if $C{\triangleright}D{\in}\Lambda$ then $C{\not\in}\Delta$, or $D{\in}\Delta$, $\mathbf{(iii)}$~$A{\in}\Lambda$ and $\mathbf{(iv)}$~$B{\not\in}\Lambda$.
%
%
\end{proposition}
\begin{proof}\footnote{The reader may easily see how $(\modalaxiom1)$, $(\modalaxiom2)$, $(\modalaxiom3)$, $(\modalaxiom5)$ and $(\rulenecessitation)$ are used in this proof.}
Suppose $\L$ contains $\L_{\partition}$.
Let $\Gamma$ be a $\L$-theory and $A,B$ be formulas.
Suppose $A{\triangleright}B{\not\in}\Gamma$.
Let ${\mathcal S}{=}\{\Delta$: $\Delta$ is a $\L$-theory such that $\Gamma{\subseteq}\Delta$ and $A{\triangleright}B{\not\in}\Delta\}$.
Since $A{\triangleright}B{\not\in}\Gamma$, therefore $\Gamma{\in}{\mathcal S}$.
Hence, ${\mathcal S}$ is nonempty.
Moreover, obviously, for all nonempty chains $(\Delta_{i})_{i{\in}I}$ of elements of ${\mathcal S}$, $\bigcup\{\Delta_{i}$: $i{\in}I\}$ is an element of ${\mathcal S}$.
Thus, by Kuratowski-Zorn Lemma, ${\mathcal S}$ possesses a maximal element $\Delta$.
Consequently, $\Delta$ is a $\L$-theory such that $\Gamma{\subseteq}\Delta$ and $A{\triangleright}B{\not\in}\Delta$.
Imitating the standard results about theories in intuitionistic propositional logic, the reader may easily demonstrate that $\Delta$ is proper and prime.
Notice that using $(\modalaxiom1)$, $B{\triangleright}B{\in}\Delta$.
Let ${\mathcal T}{=}\{(\Lambda,\Theta)$: $\Lambda$ is a $\L$-theory and $\Theta$ is a set of formulas such that for all $C{\in}\Lambda$, for all $n{\in}\N$ and for all $D_{1},\ldots,D_{n}{\in}\Theta$, $C{\triangleright}(D_{1}{\vee}\ldots(D_{n}{\vee}{\bot})\ldots){\not\in}\Delta$, for all $C{\in}\Delta$ and for all $D{\in}\Fo{\setminus}\Delta$, $C{\triangleright}D{\in}\Theta$, $A{\in}\Lambda$ and $B{\in}\Theta\}$.
\\
{\bf We claim that $(\L{+}A,\{B\}{\cup}\{C{\triangleright}D$: $C{\in}\Delta$ and $D{\in}\Fo{\setminus}\Delta\}){\in}{\mathcal T}$.}
If not, there exists $E{\in}\L{+}A$, there exists $n{\in}\N$ and there exists $F_{1},\ldots,F_{n}{\in}\{B\}{\cup}\{C{\triangleright}D$: $C{\in}\Delta$ and $D{\in}\Fo{\setminus}\Delta\}$ such that $E{\triangleright}(F_{1}{\vee}\ldots(F_{n}{\vee}{\bot})\ldots){\in}\Delta$.
Hence, $A{\rightarrow}E{\in}\L$.
Moreover,
\linebreak
there exists $n^{\prime}{\in}\N$, there exists $C_{1},\ldots,C_{n^{\prime}}{\in}\Delta$ and there exists $D_{1},\ldots,D_{n^{\prime}}{\in}\Fo{\setminus}\Delta$ such that $(F_{1}{\vee}\ldots(F_{n}{\vee}{\bot})\ldots){\rightarrow}B{\vee}((C_{1}{\triangleright}D_{1}){\vee}\ldots((C_{n^{\prime}}{\triangleright}D_{n^{\prime}}){\vee}{\bot})\ldots){\in}\L$.
Thus,
\linebreak
using $(\rulenecessitation)$, $(E{\triangleright}(F_{1}{\vee}\ldots(F_{n}{\vee}{\bot})\ldots)){\rightarrow}(A{\triangleright}(F_{1}{\vee}\ldots(F_{n}{\vee}{\bot})\ldots)){\in}\L$.
Moreover, $(C_{1}{\wedge}\ldots(C_{n^{\prime}}{\wedge}{\top})\ldots){\in}\Delta$ and $(D_{1}{\vee}\ldots(D_{n^{\prime}}{\vee}{\bot})\ldots){\not\in}\Delta$.
Consequently, by the maximality of $\Delta$ in ${\mathcal S}$, $\Delta{+}(D_{1}{\vee}\ldots(D_{n^{\prime}}{\vee}{\bot})\ldots){\not\in}{\mathcal S}$.
Hence, $A{\triangleright}B{\in}\Delta{+}(D_{1}{\vee}\ldots(D_{n^{\prime}}{\vee}
$\linebreak$
{\bot})\ldots)$.
Thus, $(D_{1}{\vee}\ldots(D_{n^{\prime}}{\vee}{\bot})\ldots){\rightarrow}(A{\triangleright}B){\in}\Delta$.
Since for all $k{\in}(n^{\prime})$, $(C_{1}{\wedge}\ldots
$\linebreak$
(C_{n^{\prime}}{\wedge}{\top})\ldots){\rightarrow}C_{k}{\in}\L$ and $D_{k}{\rightarrow}(D_{1}{\vee}\ldots(D_{n^{\prime}}{\vee}{\bot})\ldots){\in}\L$, therefore using $(\rulenecessitation)$, for all $k{\in}(n^{\prime})$, $(C_{k}{\triangleright}D_{k}){\rightarrow}((C_{1}{\wedge}\ldots(C_{n^{\prime}}{\wedge}{\top})\ldots){\triangleright}(D_{1}{\vee}\ldots(D_{n^{\prime}}{\vee}{\bot})\ldots)){\in}\L$.
Since $(F_{1}{\vee}\ldots(F_{n}{\vee}{\bot})\ldots){\rightarrow}B{\vee}((C_{1}{\triangleright}D_{1}){\vee}\ldots((C_{n^{\prime}}{\triangleright}D_{n^{\prime}}){\vee}{\bot})\ldots){\in}\L$, therefore $(F_{1}{\vee}
$\linebreak$
\ldots(F_{n}{\vee}{\bot})\ldots){\rightarrow}B{\vee}((C_{1}{\wedge}\ldots(C_{n^{\prime}}{\wedge}{\top})\ldots){\triangleright}(D_{1}{\vee}\ldots(D_{n^{\prime}}{\vee}{\bot})\ldots)){\in}\L$.
As a
\linebreak
result, using $(\rulenecessitation)$, $(A{\triangleright}(F_{1}{\vee}\ldots(F_{n}{\vee}{\bot})\ldots)){\rightarrow}(A{\triangleright}B{\vee}((C_{1}{\wedge}\ldots(C_{n^{\prime}}{\wedge}{\top})\ldots){\triangleright}
$\linebreak$
(D_{1}{\vee}\ldots(D_{n^{\prime}}{\vee}{\bot})\ldots))){\in}\L$.
Since $E{\triangleright}(F_{1}{\vee}\ldots(F_{n}{\vee}{\bot})\ldots){\in}\Delta$ and $(E{\triangleright}(F_{1}{\vee}\ldots(F_{n}
$\linebreak$
{\vee}{\bot})\ldots)){\rightarrow}(A{\triangleright}(F_{1}{\vee}\ldots(F_{n}{\vee}{\bot})\ldots)){\in}\L$, therefore $A{\triangleright}B{\vee}((C_{1}{\wedge}\ldots(C_{n^{\prime}}{\wedge}{\top})\ldots){\triangleright}
$\linebreak$
(D_{1}{\vee}\ldots(D_{n^{\prime}}{\vee}{\bot})\ldots)){\in}\Delta$.
Hence, using $(\modalaxiom5)$, $((D_{1}{\vee}\ldots(D_{n^{\prime}}{\vee}{\bot})\ldots){\rightarrow}(A{\triangleright}
$\linebreak$
B)){\rightarrow}((C_{1}{\wedge}\ldots(C_{n^{\prime}}{\wedge}{\top})\ldots){\rightarrow}(A{\triangleright}B)){\in}\Delta$.
Since $(D_{1}{\vee}\ldots(D_{n^{\prime}}{\vee}{\bot})\ldots){\rightarrow}(A{\triangleright}B){\in}
$\linebreak$
\Delta$, therefore $(C_{1}{\wedge}\ldots(C_{n^{\prime}}{\wedge}{\top})\ldots){\rightarrow}(A{\triangleright}B){\in}\Delta$.
Since $(C_{1}{\wedge}\ldots(C_{n^{\prime}}{\wedge}{\top})\ldots){\in}\Delta$,
\linebreak
therefore $A{\triangleright}B{\in}\Delta$: a contradiction.
Therefore $(\L{+}A,\{B\}{\cup}\{C{\triangleright}D$: $C{\in}\Delta$ and $D{\in}
$\linebreak$
\Fo{\setminus}\Delta\}){\in}{\mathcal T}$.
Thus, ${\mathcal T}$ is nonempty.
Moreover, obviously, for all nonempty chains $(\Lambda_{i},\Theta_{i})_{i{\in}I}$ of elements of ${\mathcal T}$, $(\bigcup\{\Lambda_{i}$: $i{\in}I\},\bigcup\{\Theta{i}$: $i{\in}I\})$ is an element of ${\mathcal T}$.
Consequently, by Kuratowski-Zorn Lemma, ${\mathcal T}$ possesses a maximal element $(\Lambda,\Theta)$.
Hence, $\Lambda$ is a $\L$-theory and $\Theta$ is a set of formulas such that
\begin{description}
\item[$({\ast})$] for all $C{\in}\Lambda$, for all $n{\in}\N$ and for all $D_{1},\ldots,D_{n}{\in}\Theta$, $C{\triangleright}(D_{1}{\vee}\ldots(D_{n}{\vee}{\bot})\ldots){\not\in}
$\linebreak$
\Delta$,
\item[$({\ast\ast})$] for all $C{\in}\Delta$ and for all $D{\in}\Fo{\setminus}\Delta$, $C{\triangleright}D{\in}\Theta$,
\item[$({\ast\ast\ast})$] $A{\in}\Lambda$ and $({\ast}{\ast}{\ast}{\ast})$~$B{\in}\Theta$.
\end{description}
{\bf We claim that $\Lambda$ is proper.}
If not, $\Lambda{=}\Fo$.
Thus, $B{\in}\Lambda$.
Since~$({\ast})$ and~$({\ast}{\ast}{\ast}{\ast})$, therefore $B{\triangleright}B{\not\in}\Delta$: a contradiction.
Therefore $\Lambda$ is proper.
\\
{\bf We claim that $\Lambda$ is prime.}
If not, there exists formulas $G,H$ such that $G{\vee}H{\in}\Lambda$, $G{\not\in}\Lambda$ and $H{\not\in}\Lambda$.
Consequently, by the maximality of $(\Lambda,\Theta)$ in ${\mathcal T}$, $(\Lambda{+}G,\Theta){\not\in}{\mathcal T}$ and $(\Lambda{+}H,
$\linebreak$
\Theta){\not\in}{\mathcal T}$.
Hence, there exists $C^{\prime}{\in}\Lambda{+}G$, there exists $n^{\prime}{\in}\N$ and there exists $D^{\prime}_{1},\ldots,D^{\prime}_{n^{\prime}}{\in}
$\linebreak$
\Theta$ such that $C^{\prime}{\triangleright}(D^{\prime}_{1}{\vee}\ldots(D^{\prime}_{n^{\prime}}{\vee}{\bot})\ldots){\in}\Delta$ and there exists $C^{\prime\prime}{\in}\Lambda{+}H$, there exists $n^{\prime\prime}{\in}\N$ and there exists $D^{\prime\prime}_{1},\ldots,D^{\prime\prime}_{n^{\prime\prime}}{\in}\Theta$ such that $C^{\prime\prime}{\triangleright}(D^{\prime\prime}_{1}{\vee}\ldots(D^{\prime\prime}_{n^{\prime\prime}}{\vee}{\bot})\ldots){\in}\Delta$.
Thus, $G{\rightarrow}C^{\prime}{\in}\Lambda$ and $H{\rightarrow}C^{\prime\prime}{\in}\Lambda$.
Consequently, $G{\vee}H{\rightarrow}C^{\prime}{\vee}C^{\prime\prime}{\in}\Lambda$.
Since $G{\vee}H{\in}\Lambda$, therefore $C^{\prime}{\vee}C^{\prime\prime}{\in}\Lambda$.
Since $(D^{\prime}_{1}{\vee}\ldots(D^{\prime}_{n^{\prime}}{\vee}{\bot})\ldots){\rightarrow}(D^{\prime}_{1}{\vee}\ldots(D^{\prime}_{n^{\prime}}{\vee}(D^{\prime\prime}_{1}{\vee}\ldots(D^{\prime\prime}_{n^{\prime\prime}}
$\linebreak$
{\vee}{\bot})\ldots))\ldots){\in}\L$ and $(D^{\prime\prime}_{1}{\vee}\ldots(D^{\prime\prime}_{n^{\prime\prime}}{\vee}{\bot})\ldots){\rightarrow}(D^{\prime}_{1}{\vee}\ldots(D^{\prime}_{n^{\prime}}{\vee}(D^{\prime\prime}_{1}{\vee}\ldots(D^{\prime\prime}_{n^{\prime\prime}}{\vee}
$\linebreak$
{\bot})\ldots))\ldots){\in}\L$, therefore using $(\rulenecessitation)$, $(C^{\prime}{\triangleright}(D^{\prime}_{1}{\vee}\ldots(D^{\prime}_{n^{\prime}}{\vee}{\bot})\ldots)){\rightarrow}(C^{\prime}{\triangleright}(D^{\prime}_{1}{\vee}
$\linebreak$
\ldots(D^{\prime}_{n^{\prime}}{\vee}(D^{\prime\prime}_{1}{\vee}\ldots(D^{\prime\prime}_{n^{\prime\prime}}{\vee}{\bot})\ldots))\ldots)){\in}\L$ and $(C^{\prime\prime}{\triangleright}(D^{\prime\prime}_{1}{\vee}\ldots(D^{\prime\prime}_{n^{\prime\prime}}{\vee}{\bot})\ldots)){\rightarrow}(C^{\prime\prime}
$\linebreak$
{\triangleright}(D^{\prime}_{1}{\vee}\ldots(D^{\prime}_{n^{\prime}}{\vee}(D^{\prime\prime}_{1}{\vee}\ldots(D^{\prime\prime}_{n^{\prime\prime}}{\vee}{\bot})\ldots))\ldots)){\in}\L$.
Since $C^{\prime}{\triangleright}(D^{\prime}_{1}{\vee}\ldots(D^{\prime}_{n^{\prime}}{\vee}{\bot})\ldots)
$\linebreak$
{\in}\Delta$ and $C^{\prime\prime}{\triangleright}(D^{\prime\prime}_{1}{\vee}\ldots(D^{\prime\prime}_{n^{\prime\prime}}{\vee}{\bot})\ldots){\in}\Delta$, therefore $C^{\prime}{\triangleright}(D^{\prime}_{1}{\vee}\ldots(D^{\prime}_{n^{\prime}}{\vee}(D^{\prime\prime}_{1}{\vee}\ldots
$\linebreak$
(D^{\prime\prime}_{n^{\prime\prime}}{\vee}{\bot})\ldots))\ldots){\in}\Delta$ and $C^{\prime\prime}{\triangleright}(D^{\prime}_{1}{\vee}\ldots(D^{\prime}_{n^{\prime}}{\vee}(D^{\prime\prime}_{1}{\vee}\ldots(D^{\prime\prime}_{n^{\prime\prime}}{\vee}{\bot})\ldots))\ldots){\in}\Delta$.
\linebreak
Hence, using $(\modalaxiom2)$, $C^{\prime}{\vee}C^{\prime\prime}{\triangleright}(D^{\prime}_{1}{\vee}\ldots(D^{\prime}_{n^{\prime}}{\vee}(D^{\prime\prime}_{1}{\vee}\ldots(D^{\prime\prime}_{n^{\prime\prime}}{\vee}{\bot})\ldots))\ldots){\in}\Delta$.
Sin\-ce~$({\ast})$, $D^{\prime}_{1},\ldots,D^{\prime}_{n^{\prime}}{\in}\Theta$, $D^{\prime\prime}_{1},\ldots,D^{\prime\prime}_{n^{\prime\prime}}{\in}\Theta$ and $C^{\prime}{\vee}C^{\prime\prime}{\in}\Lambda$, therefore $C^{\prime}{\vee}C^{\prime\prime}{\triangleright}(D^{\prime}_{1}{\vee}\ldots
$\linebreak$
(D^{\prime}_{n^{\prime}}{\vee}(D^{\prime\prime}_{1}{\vee}\ldots(D^{\prime\prime}_{n^{\prime\prime}}{\vee}{\bot})\ldots))\ldots){\not\in}\Delta$: a contradiction.
Therefore $\Lambda$ is prime.
\\
{\bf We claim that for all formulas $C,D$, if $C{\triangleright}D{\in}\Delta$ then $C{\not\in}\Lambda$, or $D{\in}\Lambda$.}
If not, there exists formulas $C,D$ such that $C{\triangleright}D{\in}\Delta$, $C{\in}\Lambda$ and $D{\not\in}\Lambda$.
Thus, by the maximality of $(\Lambda,\Theta)$ in ${\mathcal T}$, $(\Lambda{+}D,\Theta){\not\in}{\mathcal T}$.
Consequently, there exists $C^{\prime}{\in}\Lambda{+}D$, there exists $n^{\prime}{\in}\N$ and there exists $D^{\prime}_{1},\ldots,D^{\prime}_{n^{\prime}}{\in}\Theta$ such that $C^{\prime}{\triangleright}(D^{\prime}_{1}{\vee}\ldots(D^{\prime}_{n^{\prime}}{\vee}{\bot})\ldots){\in}\Delta$.
Hence, $D{\rightarrow}
$\linebreak$
C^{\prime}{\in}\Lambda$.
Since $C{\in}\Lambda$, therefore $C{\wedge}(D{\rightarrow}C^{\prime}){\in}\Lambda$.
Since $({\ast})$ and $D^{\prime}_{1},\ldots,D^{\prime}_{n^{\prime}}{\in}\Theta$, therefore $C{\wedge}(D{\rightarrow}C^{\prime}){\triangleright}(D^{\prime}_{1}{\vee}\ldots(D^{\prime}_{n^{\prime}}{\vee}{\bot})\ldots){\not\in}\Delta$.
Since $C{\triangleright}D{\in}\Delta$ and $C^{\prime}{\triangleright}(D^{\prime}_{1}{\vee}\ldots(D^{\prime}_{n^{\prime}}
$\linebreak$
{\vee}{\bot})\ldots){\in}\Delta$, therefore using $(\modalaxiom3)$, $C{\wedge}(D{\rightarrow}C^{\prime}){\triangleright}(D^{\prime}_{1}{\vee}\ldots(D^{\prime}_{n^{\prime}}{\vee}{\bot})\ldots){\in}\Delta$: a
\linebreak
contradiction.
Therefore for all formulas $C,D$, if $C{\triangleright}D{\in}\Delta$ then $C{\not\in}\Lambda$, or $D{\in}\Lambda$.
\\
{\bf We claim that for all formulas $C,D$, if $C{\triangleright}D{\in}\Lambda$ then $C{\not\in}\Delta$, or $D{\in}\Delta$.}
If not, there exists formulas $C,D$ such that $C{\triangleright}D{\in}\Lambda$, $C{\in}\Delta$ and $D{\not\in}\Delta$.
Notice that using $(\modalaxiom1)$, $(C{\triangleright}D){\triangleright}(C{\triangleright}D){\in}\Delta$.
Since $({\ast}{\ast})$, $C{\in}\Delta$ and $D{\not\in}\Delta$, therefore $C{\triangleright}D{\in}\Theta$.
Since $({\ast})$ and $C{\triangleright}D{\in}\Lambda$, therefore $(C{\triangleright}D){\triangleright}(C{\triangleright}D){\not\in}\Delta$: a contradiction.
Therefore for all formulas $C,
$\linebreak$
D$, if $C{\triangleright}D{\in}\Lambda$ then $C{\not\in}\Delta$, or $D{\in}\Delta$.
\\
{\bf We claim that $B{\not\in}\Lambda$.}
If not, since~$({\ast})$ and~$({\ast}{\ast}{\ast}{\ast})$, therefore $B{\triangleright}B{\not\in}\Delta$: a contradiction.
Therefore $B{\not\in}\Lambda$.
\medskip
\end{proof}
\section{Completeness}\label{section:completeness}
Now, we are ready to show the completeness of $\L_{\min}$ and the completeness of $\L_{\partition}$.
With this aim, we introduce the canonical model $(W_{\min},{\leq_{\min}},{R_{\min}},V_{\min})$ for $\L_{\min}$ and the canonical model $(W_{\partition},{\leq_{\partition}},{R_{\partition}},V_{\partition})$ for $\L_{\partition}$.
With Propositions~\ref{proposition:truth:lemma:for:modal:formulas}
$\linebreak$
and~\ref{proposition:truth:lemma:for:modal:formulas:L:partition}, we show results that will play for $\L_{\min}$ and for $\L_{\partition}$ the role played by the Truth Lemma in classical modal logic~\cite[Lemma~$4.21$]{Blackburn:et:al:2001}.
\\
\\
Firstly, we consider the proof of completeness of $\L_{\min}$.
\\
\\
The {\em canonical frame of $\L_{\min}$}\/ is the structure $(W_{\min},{\leq_{\min}},{R_{\min}})$ where $W_{\min}$ is the nonempty set of all prime $\L_{\min}$-theories, $\leq_{\min}$ is the preorder on $W_{\min}$ such that for all $\Gamma,\Delta{\in}W_{\min}$, $\Gamma{\leq_{\min}}\Delta$ if and only if $\Gamma{\subseteq}\Delta$ and $R_{\min}$ is the binary relation on $W_{\min}$ such that for all $\Gamma,\Delta{\in}W_{\min}$, $\Gamma{R_{\min}}\Delta$ if and only if for all formulas $A,B$, if $A{\triangleright}B{\in}\Gamma$ then $A{\not\in}\Delta$, or $B{\in}\Delta$.
%
%
%
%
The {\em canonical valuation of $\L_{\min}$}\/ is the valuation $V_{\min}\ :\ \At{\longrightarrow}\wp(W_{\min})$ on $(W_{\min},{\leq_{\min}},{R_{\min}})$ such that for all atoms $p$, $V_{\min}(p){=}\{\Gamma{\in}
$\linebreak$
W_{\min}\ :\ p{\in}\Gamma\}$.
%
%
%
%
The model $(W_{\min},{\leq_{\min}},{R_{\min}},V_{\min})$ is called {\em canonical model of $\L_{\min}$.}
\begin{proposition}\label{proposition:truth:lemma:for:modal:formulas}
Let $A$ be a modal formula.
For all $\Gamma{\in}W_{\min}$, $(W_{\min},{\leq_{\min}},{R_{\min}},
$\linebreak$
V_{\min}),\Gamma{\models}A$ if and only if $A{\in}\Gamma$.
%
%
\end{proposition}
\begin{proof}
%
%
By induction on $A$.
We only consider the ``left to right'' direction of the case $A{=}B{\triangleright}C$.
Let $\Gamma{\in}W_{\min}$.
Suppose $(W_{\min},{\leq_{\min}},{R_{\min}},V_{\min}),\Gamma{\models}B{\triangleright}C$.
For the sake of the contradiction, suppose $B{\triangleright}C{\not\in}\Gamma$.
Consequently, by Proposition~\ref{proposition:triangleright:there:exists:a:prime:theory:containing}, there exists a prime $\L$-theory $\Delta$ such that for all formulas $D,E$, if $D{\triangleright}E{\in}\Gamma$ then $D{\not\in}\Delta$, or $E{\in}\Delta$, $B{\in}\Delta$ and $C{\not\in}\Delta$.
Hence, $\Delta{\in}W_{\min}$ and $\Gamma{R_{\min}}\Delta$.
Since $B{\in}\Delta$ and $C{\not\in}\Delta$, therefore by induction hypothesis, $(W_{\min},{\leq_{\min}},{R_{\min}},V_{\min}),\Delta{\models}B$ and $(W_{\min},{\leq_{\min}},{R_{\min}},
$\linebreak$
V_{\min}),\Delta{\not\models}C$.
Since $(W_{\min},{\leq_{\min}},{R_{\min}},V_{\min}),\Gamma{\models}B{\triangleright}C$ and $\Gamma{R_{\min}}\Delta$, therefore
\linebreak$
(W_{\min},{\leq_{\min}},{R_{\min}},V_{\min}),\Delta{\models}C$: a contradiction.
\medskip
\end{proof}
%
%
%
%
%
%
%
%
%
%
%
%
%
%
%
%
%
%
%
%
%
%
%
%
%
%
%
%
%
%
%
%
\begin{proposition}[Completeness of $\L_{\min}$]\label{proposition:completeness:of:L:min}
$\L_{\min}$ is complete with respect to ${\mathcal C}_{\min}$.
\end{proposition}
\begin{proof}
For the sake of the contradiction, suppose $\L_{\min}$ is not complete with respect to ${\mathcal C}_{\min}$.
Hence, there exists a formula $A$ such that ${\mathcal C}_{\min}{\models}A$ and $A{\not\in}\L_{\min}$.
Thus, by Propositions~\ref{proposition:logic:reduction:is:always:possible} and~\ref{proposition:soundness:of:L:min}, ${\mathcal C}_{\min}{\models}\tau(A)$ and $\tau(A){\not\in}\L_{\min}$.
Consequently, by Proposition~\ref{proposition:truth:lemma:for:modal:formulas}, there exists a prime $\L_{\min}$-theory $\Gamma$ such that $(W_{\min},{\leq_{\min}},{R_{\min}},V_{\min}),\Gamma{\not\models}\tau(A)$.
Hence, ${\mathcal C}_{\min}{\not\models}\tau(A)$: a contradiction.
Therefore $\L_{\min}$ is complete with respect to ${\mathcal C}_{\min}$.
\medskip
\end{proof}
Secondly, we consider the proof of completeness of $\L_{\partition}$.
\\
\\
The {\em canonical frame of $\L_{\partition}$}\/ is the structure $(W_{\partition},{\leq_{\partition}},{R_{\partition}})$ where $W_{\partition}$ is the nonempty set of all prime $\L_{\partition}$-theories, $\leq_{\partition}$ is the preorder on $W_{\partition}$ such that for all $\Gamma,\Delta{\in}W_{\partition}$, $\Gamma{\leq_{\partition}}\Delta$ if and only if $\Gamma{\subseteq}\Delta$ and $R_{\partition}$ is the reflexive symmetric binary relation on $W_{\partition}$ such that for all $\Gamma,\Delta{\in}W_{\partition}$, $\Gamma{R_{\partition}}\Delta$ if and only if for all formulas $A,B$, if $A{\triangleright}B{\in}\Gamma$ then $A{\not\in}\Delta$, or $B{\in}\Delta$ and for all formulas $A,B$, if $A{\triangleright}B{\in}\Delta$ then $A{\not\in}\Gamma$, or $B{\in}\Gamma$.\footnote{The reader may easily demonstrate that $R_{\partition}$ is reflexive by using $(\modalaxiom4)$.
As for the symmetry of $R_{\partition}$, it immediately follows from the definitions.}
The {\em canonical valuation of $\L_{\partition}$}\/ is the valuation $V_{\partition}\ :\ \At{\longrightarrow}\wp(W_{\partition})$ on $(W_{\partition},{\leq_{\partition}},{R_{\partition}})$ such that for all atoms $p$, $V_{\partition}(p){=}\{\Gamma{\in}W_{\partition}\ :\ p{\in}\Gamma\}$.
%
%
%
%
The model $(W_{\partition},{\leq_{\partition}},{R_{\partition}},V_{\partition})$ is called {\em canonical model of $\L_{\partition}$.}
\begin{proposition}\label{proposition:truth:lemma:for:modal:formulas:L:partition}
Let $A$ be a modal formula.
For all $\Gamma{\in}W_{\partition}$, $(W_{\partition},{\leq_{\partition}},{R_{\partition}},
$\linebreak$
V_{\partition}),\Gamma{\models}A$ if and only if $A{\in}\Gamma$.
%
%
\end{proposition}
\begin{proof}
Similar to the proof of Proposition~\ref{proposition:truth:lemma:for:modal:formulas} where the use of Proposition~\ref{proposition:triangleright:there:exists:a:prime:theory:containing} is replaced by a use of Proposition~\ref{proposition:triangleright:there:exists:a:prime:theory:containing:symmetric:case}.
\medskip
\end{proof}
%
%
%
%
%
%
%
%
%
%
%
%
%
%
%
%
%
%
%
%
%
%
%
%
%
%
%
%
%
%
%
%
\begin{proposition}[Completeness of $\L_{\partition}$]\label{proposition:completeness:of:L:partition}
$\L_{\partition}$ is complete with respect to ${\mathcal C}_{\partition}$.
\end{proposition}
\begin{proof}
For the sake of the contradiction, suppose $\L_{\partition}$ is not complete with respect to ${\mathcal C}_{\partition}$.
Hence, there exists a formula $A$ such that ${\mathcal C}_{\partition}{\models}A$ and $A{\not\in}\L_{\partition}$.
Thus, by Propositions~\ref{proposition:logic:reduction:is:always:possible} and~\ref{proposition:soundness:of:L:partition}, ${\mathcal C}_{\partition}{\models}\tau(A)$ and $\tau(A){\not\in}\L_{\partition}$.
Consequently, by Proposition~\ref{proposition:truth:lemma:for:modal:formulas:L:partition}, there exists a prime $\L_{\partition}$-theory $\Gamma$ such that $(W_{\partition},{\leq_{\partition}},{R_{\partition}},V_{\partition}),\Gamma{\not\models}\tau(A)$.
As mentioned above, $R_{\partition}$ is a reflexive symmetric binary relation on $W_{\partition}$.
Unfortunately, there is no reason to believe that $R_{\partition}$ is an equivalence relation on $W_{\partition}$.
Let $(W^{\prime},{\leq^{\prime}},{R^{\prime}})$ be the frame where $W^{\prime}$ is the nonempty set of all couples of the form $(\Delta,\{\Lambda,\Theta\})$ where $\Delta,\Lambda,\Theta{\in}W_{\partition}$ are such that $\Delta{\in}\{\Lambda,\Theta\}$ and $\Lambda{R_{\partition}}\Theta$, $\leq^{\prime}$ is the preorder on $W^{\prime}$ such that for all $(\Delta,\{\Lambda,\Theta\}),(\Phi,\{\Psi,\Pi\}){\in}W^{\prime}$, $(\Delta,\{\Lambda,\Theta\}){\leq^{\prime}}(\Phi,\{\Psi,
$\linebreak$
\Pi\})$ if and only if $\Delta{\leq_{\partition}}\Phi$ and $R^{\prime}$ is the equivalence relation on $W^{\prime}$ such that for all $(\Delta,\{\Lambda,\Theta\}),(\Phi,\{\Psi,\Pi\}){\in}W^{\prime}$, $(\Delta,\{\Lambda,\Theta\}){R^{\prime}}(\Phi,\{\Psi,\Pi\})$ if and only if $\{\Lambda,\Theta\}{=}\{\Psi,
$\linebreak$
\Pi\}$.
%
%
Let $V^{\prime}\ :\ \At{\longrightarrow}\wp(W^{\prime})$ be the valuation on $(W^{\prime},{\leq^{\prime}},{R^{\prime}})$ such that for all atoms $p$, $V^{\prime}(p){=}\{(\Delta,\{\Lambda,\Theta\}){\in}W^{\prime}\ :\ p{\in}\Delta\}$.
%
%
%
%
%
%
\\
\\
{\bf Lemma.}
Let $B$ be a modal formula.
For all $(\Delta,\{\Lambda,\Theta\}){\in}W^{\prime}$, $(W^{\prime},{\leq^{\prime}},{R^{\prime}},V^{\prime}),(\Delta,
$\linebreak$
\{\Lambda,\Theta\}){\models}B$ if and only if $(W_{\partition},{\leq_{\partition}},{R_{\partition}},V_{\partition}),\Delta{\models}B$.
\\
\\
{\it Proof.}
By induction on $B$.
We only consider the ``left to right'' direction of the case $B{=}C{\triangleright}D$.
Let $(\Delta,\{\Lambda,\Theta\}){\in}W^{\prime}$.
Suppose $(W^{\prime},{\leq^{\prime}},{R^{\prime}},V^{\prime}),(\Delta,\{\Lambda,\Theta\}){\models}C{\triangleright}D$.
For the sake of the contradiction, suppose $(W_{\partition},{\leq_{\partition}},{R_{\partition}},V_{\partition}),\Delta{\not\models}C{\triangleright}D$.
Hence, there exists $\Phi{\in}W_{\partition}$ such that $\Delta{\leq_{\partition}}\Phi$ and there exists $\Psi{\in}W_{\partition}$ such that $\Phi{R_{\partition}}\Psi$, $(W_{\partition},{\leq_{\partition}},{R_{\partition}},V_{\partition}),\Psi{\models}C$ and $(W_{\partition},{\leq_{\partition}},{R_{\partition}},V_{\partition}),\Psi{\not\models}D$.
Obviously,
\linebreak$
(\Phi,\{\Phi,\Psi\})$ and $(\Psi,\{\Phi,\Psi\})$ are in $W^{\prime}$.
Moreover, $(\Delta,\{\Lambda,\Theta\}){\leq^{\prime}}(\Phi,\{\Phi,\Psi\})$ and $(\Phi,
$\linebreak$
\{\Phi,\Psi\}){R^{\prime}}(\Psi,\{\Phi,\Psi\})$.
Since $(W_{\partition},{\leq_{\partition}},{R_{\partition}},V_{\partition}),\Psi{\models}C$ and $(W_{\partition},{\leq_{\partition}},
\linebreak
{R_{\partition}},V_{\partition}),\Psi{\not\models}D$, therefore by induction hypothesis, $(W^{\prime},{\leq^{\prime}},{R^{\prime}},V^{\prime}),(\Psi,\{\Phi,\Psi\})
$\linebreak$
{\models}C$ and $(W^{\prime},{\leq^{\prime}},{R^{\prime}},V^{\prime}),(\Psi,\{\Phi,\Psi\}){\not\models}D$.
Since $(\Delta,\{\Lambda,\Theta\}){\leq^{\prime}}(\Phi,\{\Phi,\Psi\})$ and $(\Phi,
$\linebreak$
\{\Phi,\Psi\}){R^{\prime}}(\Psi,\{\Phi,\Psi\})$, therefore $(W^{\prime},{\leq^{\prime}},{R^{\prime}},V^{\prime}),(\Delta,\{\Lambda,\Theta\}){\not\models}C{\triangleright}D$: a contradicti\-on.
\\
\\
Since $(W_{\partition},{\leq_{\partition}},{R_{\partition}},V_{\partition}),\Gamma{\not\models}\tau(A)$, then $(W^{\prime},{\leq^{\prime}},{R^{\prime}},V^{\prime}),(\Gamma,\{\Gamma\}){\not\models}\tau(A)$.
\linebreak
Thus, ${\mathcal C}_{\partition}{\not\models}\tau(A)$: a contradiction.
Therefore $\L_{\partition}$ is complete with respect to ${\mathcal C}_{\partition}$.
\medskip
\end{proof}
And in the end, we compare $\L_{\min}$ and $\L_{\partition}$ to the intuitionistic public announcement logics studied in~\cite{Ma:Palmigiano:Sadrzadeh:2014,Maffezioli:Negri:2011}.
\begin{proposition}\label{proposition:L:partition:is:strictly:reducible:to:IPAL}
\begin{description}
\item[$\mathbf{(1)}$] $\L_{\min}{\cap}\Fo_{\direct}$ is strictly contained in $\IPAL$ interpreted over the class of all backward and forward confluent models,
\item[$\mathbf{(2)}$] $\L_{\partition}{\cap}\Fo_{\direct}$ is strictly contained in $\IPAL$ interpreted over the class of all backward and forward confluent models where the modal accessibility relation is an equivalence relation.
\end{description}
\end{proposition}
\begin{proof}
$\mathbf{(1)}$~It suffices to notice that the instance ${\lbrack p\rbrack}{\square}r{\leftrightarrow}(p{\rightarrow}{\square}{\lbrack p\rbrack}r)$ of the reduction
\linebreak
axiom for ``necessity'' ${\lbrack A\rbrack}{\square}C{\leftrightarrow}(A{\rightarrow}{\square}{\lbrack A\rbrack}C)$ considered in the literature of $\IPAL$ is not valid in the class of all models.
Indeed, the reader may easily verify that $0{\models}{\lbrack p\rbrack}{\square}r$ and $0{\not\models}p{\rightarrow}{\square}{\lbrack p\rbrack}r$ in the model $(W,{\leq},{R},V)$ such that $W=\{0,1,2\}$, $\leq{=}\{(0,0),(1,
$\linebreak$
1),(1,2),(2,2)\}$, $R{=}\{(0,0),(0,1),(1,0),(1,1),(2,2)\}$, $V(p){=}\{0,2\}$ and $V(r){=}
$\linebreak$
\{0\}$.
\\
\\
$\mathbf{(2)}$~Similar to the proof of Item~$\mathbf{(1)}$.
\medskip
\end{proof}
\section{Conclusion}
Much remains to be done.
\\
\\
First, recall that direct modal formulas can also be considered as $\lozenge$-free formulas of the language of intuitionistic modal logics.
In particular, the restriction of $\L_{\min}$ to the set of all its direct modal formulas is equal to the so-called ``intuitionistic modal logic $\K$ without $\lozenge$'' considered by Simpson~\cite[Chapter~$3$]{Simpson:1994}, an intuitionistic modal logic called $\iK$ by Almeida {\em et al.}\/~\cite{Almeida:et:al:2026} and van der Giessen~\cite{VanDerGiessen:2023}.
Therefore the restriction of $\L_{\min}$ to the set of all its modal formulas deserves to be studied for its own sake.
In doing so, one should not forget to compare the new binary operator $\triangleright$ of ``relativized necessity'' and the connective of strict implication firstly developed in~\cite{Lewis:Langford:1932} and recently studied in~\cite{deGroot:et:al:2026,Litak:Visser:2018}.
\\
\\
Second, recall that the dual $\lozenge$ of $\square$ introduced by P\v{r}enosil~\cite{Prenosil:2014} and developed by Balbiani and Gencer~\cite{Balbiani:Gencer:IGPL:2025,Balbiani:Gencer:Studia:Logica:2025} is such that for all formulas $A$, given a model $(W,{\leq},{R},V)$ and $s{\in}W$, $(W,{\leq},{R},V),s{\models}{\lozenge}A$ if and only if there exists $t{\in}W$ such that $t{\leq}s$ and there exists $u{\in}W$ such that $t{R}u$ and $(W,{\leq},{R},V),u{\models}A$.
Therefore the extension of $\L_{\min}$ by the dual $\triangleleft$ of $\triangleright$ deserves to be studied for its own sake.
It could be such that for all formulas $A,B$, given a model $(W,{\leq},{R},V)$ and $s{\in}W$, $(W,{\leq},{R},V),s{\models}A{\triangleleft}B$ if and only if there exists $t{\in}W$ such that $t{\leq}s$ and there exists $u{\in}W$ such that $t{R}u$, $(W,{\leq},{R},V),u{\not\models}A$ and $(W,{\leq},{R},V),u{\models}B$.
\\
\\
Third, in order to evaluate the computability of $\L_{\min}$ and the computability of $\L_{\partition}$, one may design tableau calculi or sequent calculi for these intuitionistic public announcement logics, following the lines of reasoning adopted in~\cite{Balbiani:et:al:2010,Lerouvillois:Poggiolesi:2026} with respect to the classical setting and in~\cite{Balbiani:Galmiche:2016,Dalmonte:et:al:2021,Nomura:Sano:Tojo:2015} with respect to the intuitionistic setting.
Calculi for the restrictions of $\L_{\min}$ and $\L_{\partition}$ to the sets of all their modal formulas deserve to be studied for their own sake as well.
\\
\\
Fourth, one may develop the multi-agent version with common knowledge, or distributed knowledge of $\L_{\min}$ and $\L_{\partition}$.
In this multi-agent version, one has to determine which reduction axioms make it possible to associate with each formula a semantically equivalent formula containing no public announcement operators.
See~\cite{Jager:Marti:2016,Murai:Sano:2022,Su:et:al:2021,Zenger:2026} for various multi-agent version with common knowledge, or distributed knowledge of intuitionistic modal logics.
\\
\\
Fifth, in a multi-agent version with common knowledge, or distributed knowledge of $\L_{\min}$ and $\L_{\partition}$, a successful announcement is a formula that always becomes known after being announced.
Many dynamic phenomena can be explained by analyzing successful announcements~\cite{vanDitmarsch:Kooi:2006}.
As a result, one may investigate the syntactic characterization of the successful announcements within the context of intuitionistic public announcement logic.
%
%
%
%
%
%
%
%
\section*{Acknowledgements}
Special acknowledgement is granted to our colleagues of the Toulouse Institute of Computer Science Research for many stimulating discussions about the subject of this article.
We also thank the reviewers for their feedback: their insightful suggestions were essential in improving the readability of an earlier version of this work.
\bibliographystyle{named}
\end{document}